\documentclass[sigconf]{acmart}
\AtBeginDocument{%
  }

\usepackage{balance}
\usepackage{todonotes}
\usepackage{colortbl}
\usepackage{physics}
\usepackage{graphicx}
\usepackage{algorithm}
\usepackage{algorithmic}
\usepackage{multirow}
\usepackage{makecell}
\copyrightyear{2025}
\acmYear{2025}
\setcopyright{acmlicensed}
\acmDOI{10.1145/3746027.3754568}
\acmConference[MM '25] {Proceedings of the 33rd ACM International Conference on Multimedia}{October 27--31, 2025}{Dublin, Ireland.}
\acmBooktitle{Proceedings of the 33rd ACM International Conference on Multimedia (MM '25), October 27--31, 2025, Dublin, Ireland}
\acmISBN{979-8-4007-2035-2/2025/10}

\begin{document}

\title{Multi-level SSL Feature Gating for Audio Deepfake Detection}

\author{Hoan My Tran}
\orcid{0009-0006-6818-6157}
\affiliation{%
  \institution{Univ Rennes, IRISA, CNRS}
  \city{Lannion}
  \country{France}
}
\email{hoan.tran@irisa.fr}

\author{Damien Lolive}
\orcid{0000-0002-1110-5444}
\affiliation{%
  \institution{Univ Bretagne Sud, IRISA, CNRS}
  \city{Vannes}
  \country{France}
}
\email{damien.lolive@irisa.fr}

\author{Aghilas Sini}
\orcid{0000-0001-6785-6936}
\affiliation{%
  \institution{Univ Le Mans, LIUM}
  \city{Le Mans}
  \country{France}
}
\email{aghilas.sini@univ-lemans.fr}

\author{Arnaud Delhay}
\orcid{0000-0001-6795-7999}
\affiliation{%
  \institution{Univ Rennes, IRISA, CNRS}
  \city{Lannion}
  \country{France}
}
\email{arnaud.delhay@irisa.fr}

\author{Pierre-François Marteau}
\orcid{0000-0002-3963-8795}
\affiliation{%
  \institution{Univ Bretagne Sud, IRISA, CNRS}
  \city{Vannes}
  \country{France}
}
\email{pierre-francois.marteau@irisa.fr}

\author{David Guennec}
\orcid{0009-0006-3265-6321}
\affiliation{%
  \institution{Univ Rennes, IRISA, CNRS}
  \city{Lannion}
  \country{France}
}
\email{david.guennec@irisa.fr}

\renewcommand{\shortauthors}{Hoan My Tran et al.}

\begin{abstract}
  Recent advancements in generative AI, particularly in speech synthesis, have enabled the generation of highly natural-sounding synthetic speech that closely mimics human voices. While these innovations hold promise for applications like assistive technologies, they also pose significant risks, including misuse for fraudulent activities, identity theft, and security threats. Current research on spoofing detection countermeasures remains limited by generalization to unseen deepfake attacks and languages. To address this, we propose a gating mechanism extracting relevant feature from the speech foundation XLS{\textendash}R model as a front{\textendash}end feature extractor. For downstream back{\textendash}end classifier, we employ Multi{\textendash}kernel gated Convolution (MultiConv) to capture both local and global speech artifacts. Additionally, we introduce Centered Kernel Alignment (CKA) as a similarity metric to enforce diversity in learned features across different MultiConv layers. By integrating CKA with our gating mechanism, we hypothesize that each component helps improving the learning of distinct synthetic speech patterns. Experimental results demonstrate that our approach achieves state{\textendash}of{\textendash}the{\textendash}art performance on in{\textendash}domain benchmarks while generalizing robustly to out{\textendash}of{\textendash}domain datasets, including multilingual speech samples. This underscores its potential as a versatile solution for detecting evolving speech deepfake threats.
\end{abstract}



\maketitle

\section{Introduction}
Text-To-Speech (TTS) and Voice Conversion (VC) have enabled the synthesis of highly realistic speech through deep neural networks. However, these technologies are increasingly misused for political manipulation, social media disinformation, and economic fraud, necessitating robust defenses for Automatic Speaker Verification (ASV) systems. To address this, research in anti-spoofing and Synthetic Speech Detection (SSD) has intensified, with the ASVspoof challenge series \cite{wu15e_interspeech, kinnunen17_interspeech, todisco2019asvspoof, yamagishi21_asvspoof, wang24_asvspoof} emerging as the benchmark for developing CounterMeasure (CM) systems.

Traditional CM systems rely on a front-end feature extractor (e.g., MFCC, CQCC) \cite{10.5120/20312-2362, TODISCO2017516}, paired with a back-end classifier to distinguish spoofed from bona fide speech. Recent work has shifted towards Self-Supervised Learning (SSL) features extracted from foundation speech models, which combine Convolutional Neural Network (CNN) layers with Transformer encoders \cite{NIPS2017_3f5ee243} based on Multi-Layer Perceptron (MLP) backbones. For different downstream tasks, the Conformer architecture \cite{gulati20_interspeech} was proposed as an improvement to Transformers, combining CNNs and Transformers to model both local and global dependencies. This has shown effectiveness in Automatic Speech Recognition (ASR) as well as in SSD \cite{rosello23_interspeech, truong24b_interspeech}. While Transformers and Conformers leverage self-attention mechanisms, alternative architectures like gated MLP (gMLP) \cite{liu2021pay} use trainable gating mechanisms to filter selective features, demonstrating effectiveness in localizing partially spoofed audio \cite{zhang23v_interspeech, zhang24j_interspeech}. Similarly, MultiConv \cite{prabhu24_interspeech} fuses multiple CNN kernels to capture both local and global speech patterns. While successful in tasks like ASR, it has yet to be explored for SSD. Additionally, using feature gating helps reduce the computation cost \cite{NEURIPS2019_68b1fbe7}, and the overfitting in data with highly redundant features \cite{DENG2021101182}.

Speech foundation models with deep architectures enhance pattern discovery through hierarchical representations, where successive layers encode correlations between acoustic, paralinguistic, and linguistic features \cite{9688093, 10096149}. However, layer-wise analyses reveal redundancy \cite{doi:10.1073/pnas.2015509117}, with adjacent Transformer layers often learning overlapping correlations, which limits feature diversity. In contrast, gMLP and MultiConv architectures utilize gating mechanisms to extract sparse, selective features. While these approaches share foundational principles in representation learning, their inter-layer differences remain underexplored. Several studies have used similarity metrics to reduce redundancy, thereby minimizing the number of parameters \cite{jiang2025tracing}, and increasing diversity in feature learning \cite{10658409}. However, the potential for leveraging dissimilarity across layers to enhance the detection of diverse spoofed speech artifacts in SSD has yet to be fully explored.

In this work, we hypothesize that hierarchical gating mechanisms within MultiConv layers can learn complementary discriminative features for SSD. Our main contributions are summarized as follows:

\begin{itemize}
    \item We aggregate XLS-R hidden features using Swish-Gated Linear Unit (SwiGLU) activation \cite{shazeer2020glu} for dynamic self-gating, thereby enhancing artifact-sensitive feature selection.
    \item We stack gated MultiConv layers to model layer-specific local and global dependencies, utilizing these to improve deepfake detection.
    \item We employ CKA as a loss function to minimize inter-layer redundancy within MultiConv, promoting the learning of distinct features.
    \item We evaluate the performance of our model on diverse datasets, demonstrating its ability to generalize across different language families, including Germanic, Romance, Slavic, and Sino-Tibetan.
\end{itemize}

\section{Related Work}

Recent approaches predominantly adopt a two-stage pipeline comprising a front-end feature extractor (e.g., HuBERT, WavLM, Wav2Vec 2.0, XLS-R, and MMS) \cite{10.1109/TASLP.2021.3122291, 9814838, NEURIPS2020_92d1e1eb, babu22_interspeech, JMLR:v25:23-1318} followed by a back-end classifier. These foundation models, pre-trained on large-scale datasets, extract highly relevant features and significantly improve detection performance by mitigating the limitations of training data.

Tak et al. \cite{tak22_odyssey} pioneered the use of XLS-R features paired with a graph-based end-to-end classifier (AASIST) \cite{9747766}, demonstrating robust performance under channel variations in the ASVspoof 2021 Logical Access (21LA) sub-challenge \cite{yamagishi21_asvspoof}. Subsequently, Rosello et al. \cite{rosello23_interspeech} introduced a Conformer-based architecture that leverages self-attention mechanisms to effectively model artifacts introduced in spoofed speech. Building upon this, Truong et al. \cite{truong24b_interspeech} further improved performance by integrating a Temporal-Channel Modeling (TCM) module to capture inconsistencies in synthetic speech. More recently, Xiao et al. \cite{10909468} proposed a Mamba-based classifier \cite{gu2024mamba} that replaces the self-attention mechanism and achieves strong results on both the 21LA and ASVspoof 2021 DeepFake (21DF) sub-challenges \cite{yamagishi21_asvspoof}, setting a new State-Of-The-Art (SOTA) on the out-of-domain In-The-Wild (ITW) dataset \cite{muller22_interspeech}.

Beyond detector architectures, recent efforts have focused on effectively exploiting foundation models to extract rich and meaningful features for improving SSD systems \cite{kheir2025comprehensive}. Martín-Doñas et al. \cite{9747768} explored contextualized speech representations across different Transformer layers with learnable weights to capture discriminative information. Building on this, Zhang et al. \cite{zhang2024audio} proposed a Sensitive Layer Selection (SLS) classifier to optimize layer selection for Transformer encoders. Huang et al. \cite{10888328} enhanced generalization via Latent Space Refinement (LSR) and Augmentation (LSA). Wang et al. \cite{wang2024mixture} introduced a Mixture-Of-Experts (MOE) framework that dynamically routes frozen Wav2Vec 2.0 features to specialize detectors. Jin et al. \cite{10890563} combined cross-modal spectrograms with SSL aggregation to leverage multi-scale representations. Tran et al. \cite{tran24_interspeech} improved task-specific layer selection by prioritizing WavLM layers sensitive to speaker-related objectives based on the original method from \cite{10447923}, while Pan et al. \cite{pan24c_interspeech} proposed a method to attentively merge hidden embeddings from different Transformer layers.

Gating mechanisms have also shown promising in detecting partially spoofed speech. For instance, stacking gMLP has demonstrated the ability to localize spoofed regions within utterances by learning distinctive features \cite{zhang23v_interspeech, zhang24j_interspeech}. However, their application to fully spoofed speech detection remains unexplored. Additionally, combining multiple convolution kernels within a convolutional block \cite{prabhu24_interspeech}, especially when integrated with gating can improve the modeling of local dependencies at various granularities. This approach is also more parameter-efficient and less computationally intensive than Transformer or Conformer architectures, which rely heavily on resource consuming self-attention mechanisms.

Prior research has primarily relied on English training data, such as the ASVspoof 2019 Logical Access (19LA) dataset \cite{WANG2020101114}, to train SSD systems. These systems were typically evaluated on both in-domain datasets (e.g., 21LA and 21DF) and out-of-domain datasets (e.g., ITW). More recently, several studies have begun to evaluate SSD performance on multilingual speech datasets with previously unseen attack types \cite{chetia-phukan-etal-2024-heterogeneity, marek2024audio}.

\section{Proposed Method}

\begin{figure*}[tb!]
    \centering
    \includegraphics[]{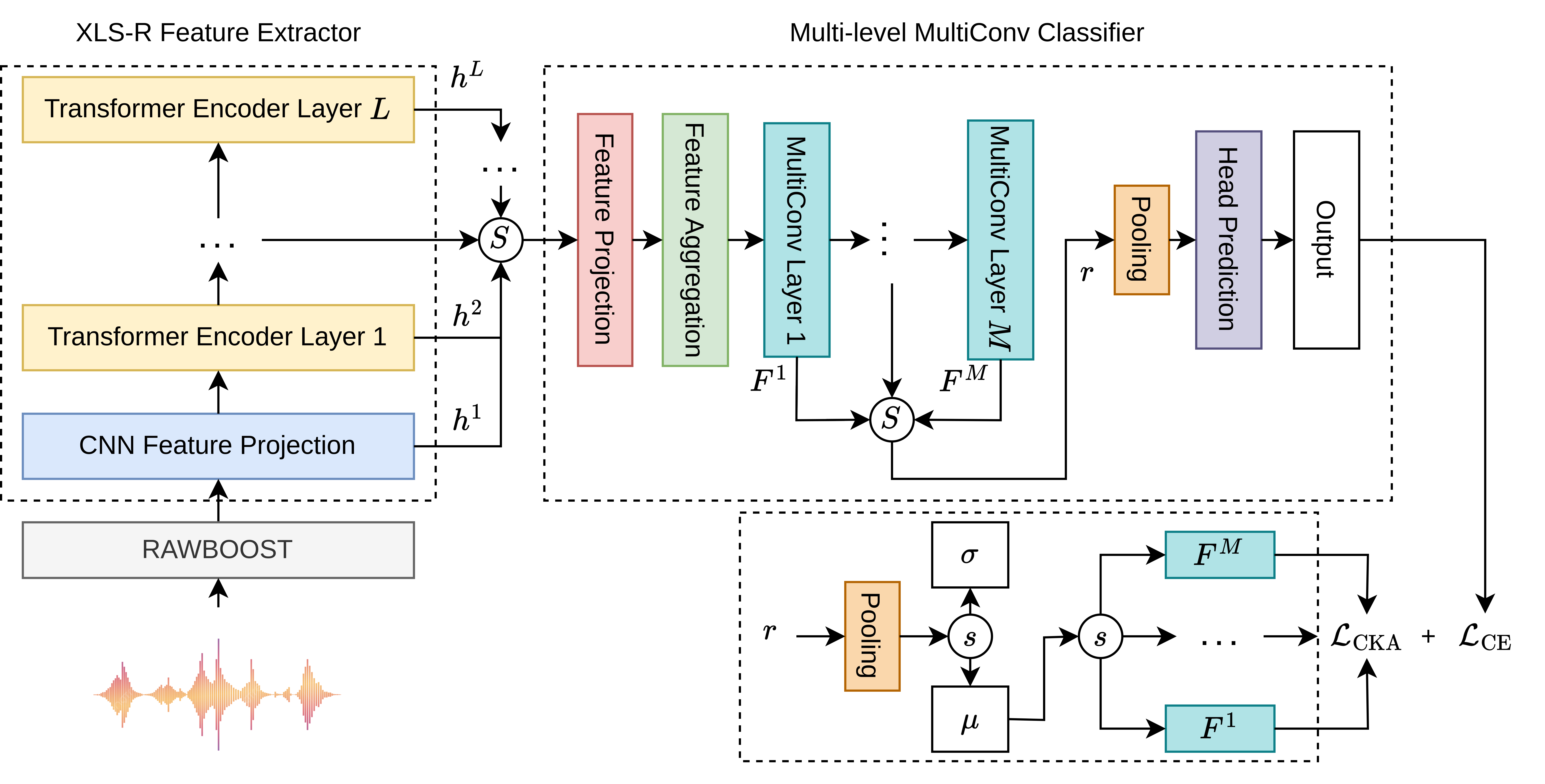}
    \caption{Overview of the proposed model. SSL features are extracted from the input waveform. Hidden states are stacked, projected to a lower dimension, aggregated, and gated. Refined features pass through stacked MultiConv blocks, pooled, and classified as bona fide or spoofed. $\mathcal{L}_{\text{CKA}}$ is used to compute dissimilarity between MultiConv outputs. \(S\) denotes the stacking operation, \(s\) represents the split operation, \(\mu\) indicates the mean, and \(\sigma\) denotes the standard deviation.}
    \label{fig:architecture}
\end{figure*}
In this section, we detail our pipeline for speech deepfake detection as shown in Figure \ref{fig:architecture}, structured into four core components. First, we introduce the XLS-R front-end as a feature extractor. Next, we describe the gating mechanism for aggregating SSL hidden features. Building on this, we present the MultiConv architecture as the back-end classifier. Finally, we integrate Multi-Head Attention Pooling (MHAP) to aggregate frame-level features, followed by a MLP classifier trained with a joint loss function combining Cross Entropy (CE) and CKA.

\subsection{XLS-R Front-End Feature Extractor}

XLS-R \cite{babu22_interspeech} is an extension of Wav2Vec2.0 \cite{NEURIPS2020_92d1e1eb} designed for cross-lingual speech representation learning using SSL techniques. The model has been trained on 436,000 hours of publicly available speech data spanning 128 languages \cite{wang-etal-2021-voxpopuli, pratap20_interspeech, ardila-etal-2020-common, valk2021slt, DBLP:conf/sltu/GalesKRR14}, enabling robust performance in various cross-lingual speech processing tasks.
The architecture of XLS-R comprises a convolutional feature encoder followed by Transformer-based context networks. The feature encoder \( f: X \mapsto Z \) consists of 7 CNN layers, which process raw audio waveforms \(X\) into latent speech representations \( z_1, \dots, z_{T} \) over \(T\) time steps using a sliding window of 25 ms with a stride of 20 ms. Inspired by BERT’s masked language modeling approach, XLS-R learns contextualized representations by randomly masking feature vectors before feeding them into the Transformer layers. The encoded speech representations \(Z\) serve as inputs to a stack of 24 Transformer layers \( g: Z \mapsto C \), which generate contextualized representations \( c_1, \dots, c_{T} \).

Inspired by \cite{9747768}, which explores the aggregation of hidden representations \( h^1, \dots, h^{L} \), we process the audio input through the SSL feature extractor, obtaining a sequence of frames of length \( T \) across \( L \) hidden states with \(D\)-dimensional space, including the feature projection layer from the final CNN layer. To construct the aggregated representation, we stack all hidden representations as:  
\begin{equation}
H = (h^1, \ldots, h^{L}) \in \mathbb{R}^{L \times T \times D}.
\end{equation}  

We then apply a projection to map \(H\) into a \( U \)-dimensional space. Inspired by the sensitive layer selection module, which enables dynamic channel-wise recalibration of feature maps \cite{zhang2024audio}, we employ the SwiGLU activation function. SwiGLU is a variant of the gated linear unit that integrates the Swish activation function into its gating mechanism, enhancing the model’s ability to capture complex relationships between input features and output representations:
\begin{equation}
    \text{SwiGLU}(H) = sigmoid(H W_1) \odot (H W_2),
\end{equation}
where \( W_1, W_2 \in \mathbb{R}^{U} \) are learnable weight matrices, and \( \odot \) denotes element-wise multiplication.
The final aggregated output, \(\mathbf{H}_{\text{agg}} \in \mathbb{R}^{T \times U}\), is computed as: 
\begin{equation}
    \mathbf{H}_{\text{agg}}(t) = \sum_{l=1}^{L} H(l, t), \quad \forall t \in \{1, \dots, T\}.
\end{equation}  

\subsection{Multi-Kernel Gated Convolution Classifier}
We employ the MultiConv module as our back-end classifier, inspired by \cite{zhang23v_interspeech, zhang24j_interspeech}. MultiConv \cite{prabhu24_interspeech}, a variant of gMLP, leverages multiple convolutional kernels in conjunction with gating mechanisms to effectively model local dependencies at various granularities. The hidden representations are first normalized, followed by an expansion of the channel dimension from \( U \) to \( d_{\text{inter}} \) using a GELU activation function. The transformed representations are then processed through the multi-kernel convolutional spatial gating unit, which integrates convolutional operations with gating mechanisms to enhance feature selection.

Following \cite{prabhu24_interspeech}, we employ the MultiConv module to capture both local and global dependencies, enhancing the modeling of frame-level discriminative features. Initially, the aggregated representation \( \mathbf{H}_{\text{agg}} \) is projected to a higher-dimensional space \( d_{\text{inter}} \) as follows:
\begin{equation}
    \hat{E} = \text{GELU}(\text{Proj}(\mathbf{H}_{\text{agg}})) \in \mathbb{R}^{T \times d_{\text{inter}}}.
\end{equation}

Next, the transformed representation \( \hat{E} \) is split into two parts, \( Z_l \) and \( Z_r \), where the dimensionality \( d' \) is defined as \( d_{\text{inter}}/2 \). A set of \( P \) convolutional operations with kernel sizes \( \{k_1, k_2, \dots, k_P\} \) is then applied to \( Z_r \) as follows:
\begin{equation}
    \begin{aligned}
        Z_l &= \hat{E}[:, :d'], \quad Z_r = \text{LN}(\hat{E}[:, d':]), \\
        V_j &= \text{Conv}_{k_j}(Z_r), \quad j = 1, 2, \dots, P,\\
        \tilde{Z}_r &= \text{Fusion}([V_1, V_2, \dots, V_P]) \in \mathbb{R}^{T \times d'},
    \end{aligned}
\end{equation}
where \( \tilde{Z}_r \) represents the fused outputs of the convolutional layers, and LN denotes Layer Normalization.

Finally, a gating mechanism is applied to integrate \( \tilde{Z}_r \) and \( Z_l \), before projecting the result back to the original dimensionality \( U \):
\begin{equation}
    F = \text{Dropout}(\text{Proj}(\tilde{Z}_r \odot Z_l)) \in \mathbb{R}^{T \times U}.
\end{equation}

To obtain the final output representation \( r \), we employ MHAP method \cite{india19_interspeech}, which divides the hidden states into \( k \) heads, each representing a sub-vector. Let \( G = (F^1, \ldots, F^{M}) \in \mathbb{R}^{T \times Q} \) denote the output of the stacked MultiConv layers where \( Q = M \times U \). The hidden state at time step \( t \) is then represented as \( G_{t} = [G_{t,1}, \ldots, G_{t,k}] \), where \( G_{t,j} \in \mathbb{R}^{Q/k} \) is the \( j \)-th sub-vector corresponding to the \( j \)-th head. Each \( j \)-th head is computed as follows:

\begin{equation}
    r_j = \sum_{t=1}^T G_{t,j}^\top \frac{\exp(G_{t,j}^\top u_j)}{\sum_{l=1}^T \exp(G_{l,j}^\top u_j)}, \quad j = 1, \dots, k.
\end{equation}

The final output representation \( r \) is obtained by concatenating the representations of all \( k \) heads, followed by the computation of the mean \( \mu \) and standard deviation \( \sigma \) of the resulting vector. This is then passed through a MLP classification head to determine whether the speech is genuine or spoofed:

\begin{equation}
    r = [r_1, \dots, r_k], \quad o = \text{MLP}([\mu(r), \sigma(r)], 2).
\end{equation}

Before passing the features to the classification stage, we stack multiple gated MultiConv layers to enhance discriminative feature learning through gating mechanisms.
The processed feature representation \(F\) is passed through \(M\) MultiConv layers to capture both low and high-level features.

\subsection{Multi-level Gating Features}

To effectively learn different level of the gating features through MultiConv layers, we employ CKA as a loss function \(\mathcal{L}_{\text{CKA}}\) to enhance the dissimilarity between each layer.
CKA has been introduced as a robust and reliable metric to measure representational similarity between features. Originally proposed by \cite{kornblith2019similarity}, CKA quantifies the alignment between two sets of neural activations and is defined as:

\begin{equation}
    \text{CKA}(K, N) = \frac{\text{HSIC}(K, N)}{\sqrt{\text{HSIC}(K, K)\text{HSIC}(N, N})},
\end{equation}

where Hilbert-Schmidt Independence Criterion (HSIC) is given by:

\begin{equation}
    \text{HSIC}(K, N) = \frac{\text{trace}(K J_m N J_m)}{(m-1)^2},
\end{equation}

where \( J_m = I_m - \frac{1}{m} 11^\top \) is the centering matrix. In linear CKA, the similarity is computed using Gram matrices \( K = S S^\top \) and \( N = Y Y^\top \), where \( S \in \mathbb{R}^{m \times p_1} \) and \( Y \in \mathbb{R}^{m \times p_2} \) represent the activation matrices of two network layers. Here, \( m \) is the number of input samples, and \( p_1, p_2 \) denote the number of neurons in each layer. Importantly, CKA is invariant to differences in layer dimensionality, meaning that layers with different numbers of neurons can still be compared. To align feature distributions across layers, we compute the \(\mathcal{L}_{\text{CKA}}\) as follows. For every pair of \(l\)-th MultiConv layer \( (p, q) \):  
\begin{equation}  
    \mathcal{L}_{\text{CKA}} = \frac{2}{M_{l}(M_{l}-1)} \sum_{p=1}^{M_{l}} \sum_{q=p}^{M_{l}} \left( \text{CKA}(p,q)\right),  
\end{equation}  
where \(M\) is the total number of MultiConv layers. We use \(\mathcal{L}_{\text{CE}}\) loss which is computed as:
\begin{equation}
    \mathcal{L}_{\text{CE}} = -\frac{1}{\hat{N}} \sum_{i=1}^{\hat{N}} \sum_{j=1}^{\hat{C}} y_{i,j} \log\left(\hat{y}_{i,j}\right),
\end{equation}
where \(\hat{C}\) is the number of classes, \(\hat{N}\) represents the number of samples in a batch, \(y\) is the true label and \(\hat{y}\) is the predicted as spoofed or bona fide speech. The final training objective is computed as:  
\begin{equation}  
    \mathcal{L}_{\text{Final}} = \mathcal{L}_{\text{CE}} + \mathcal{L}_{\text{CKA}}.
\end{equation}

\section{Experiments}
In this section, we experiment different configurations of the system designed previously. First, the datasets are introduced. Then, we present the performance metric with experimental settings. Finally, results are discussed and compared to SOTA systems. The source code will be made available on Github\footnote{\url{https://github.com/hoanmyTran/dissimilarity_deepfake_detection}}. 

\subsection{Datasets}
For training our systems, we used the ASVspoof 2019 Logical Access (19LA) training set \cite{WANG2020101114}. To assess the model's generalization to other datasets, we selected the model that achieved the best performance on the 19LA development set \cite{WANG2020101114}. Notably, the training and development sets feature different speakers. The ASVspoof 2019 dataset comprises spoofed speech generated using TTS and VC techniques, with all samples originating from the VCTK database \cite{Yamagishi2019CSTRVC}. However, the dataset consists of clean speech recordings, devoid of noise or channel variations, which may limit its applicability to real-world scenarios.

For the evaluation phase, we assess our models on the 19LA evaluation set, as well as the 21LA and 21DF evaluation sets. The 19LA evaluation set contains spoofed speech generated by 13 previously unseen algorithms that were not present in the training or development sets. 21LA extends 19LA by incorporating codec and transmission effects to better simulate real-world conditions. The dataset includes speech transmitted through real telephone systems, covering a range of codecs, transmission channels, bitrates, and sampling rates. The 21DF subset further introduces variability by applying different lossy compression techniques during audio transmission. Both bona fide and spoofed speech utterances are processed with diverse vocoders, including previously unseen ones from the Voice Conversion Challenge (VCC) 2018 \cite{lorenzotrueba18_odyssey} and VCC 2020 \cite{yi20_vccbc} challenges.

For out-of-domain evaluation, we assess our models on a diverse set of datasets, including the original version of Fake or Real (FoR) \cite{8906599}, ITW \cite{muller22_interspeech}, the Diffusion and Flow-matching-based Audio Deepfake Dataset (DFADD) \cite{10832250}, LibriSeVoc \cite{Sun_2023_CVPR}, and the DEepfake CROss-lingual (DECRO) English (D-EN) and Chinese (D-CH) \cite{10.1145/3543507.3583222} evaluation sets. We also use Multi-Language Audio Anti-Spoof (MLAAD) \cite{10650962} including English (M-EN), French (M-FR), German (M-DE), Spanish (M-ES), Italian (M-IT), Polish (M-PL), Russian (M-RU) and Ukrainian (M-UK). Additionally, we evaluate our models on the Audio Deepfake Detection 2023 \cite{DBLP:conf/dada/YiTFYWWZZZRXZGW23} dataset across Track 1.2 in both Round 1 (ADD23-R1) and Round 2 (ADD23-R2), as well as the Spanish HABLA \cite{tamayoflorez23_interspeech} dataset.

The FoR dataset is an English-language collection containing both bona fide and spoofed speech, generated by a variety of TTS algorithms sourced from open datasets, TED Talks, and YouTube videos. ITW, on the other hand, consists of bona fide audio from English-speaking celebrities and politicians, collected from publicly available sources such as social media and video streaming platforms. The DFADD dataset includes deepfake audio created using advanced diffusion and flow-matching TTS models. The LibriSeVoc dataset was specifically created to study vocoder artifacts. DECRO is designed for evaluating SDD systems in a cross-lingual context, containing fake and real audio clips in both English and Chinese. The ADD23 Track 1.2 dataset focuses on detecting fake utterances, specifically those from Track 1.1 of the ADD23 challenge. The evaluation phase is divided into two datasets, ADD23-R1 and ADD23-R2. The HABLA dataset is a Spanish language anti-spoofing corpus, representing accents from Argentina, Colombia, Peru, Venezuela, and Chile. It includes over 22,000 genuine speech samples from male and female speakers across these countries, along with 58,000 spoofed samples generated using six different speech synthesis methods. MLAAD is a multilingual speech deepfake dataset created using 54 TTS models across 23 languages sourced from M-AILABS. Statistics for all these datasets are provided in Table 1.

\begin{table}[h]
    \centering
    \caption{Statistics of datasets used in the study.}
    \resizebox{\columnwidth}{!}{%
    \begin{tabular}{l c r r c}
        \toprule
        \textbf{Dataset} & \textbf{Language} & \textbf{\# Bona fide} & \textbf{\# Spoofed} & \textbf{Attack}\\
        \midrule
\multicolumn{5}{c}{\centering \textbf{Training}} \\ 
\midrule
        19LA \cite{WANG2020101114}    & English           & 2,580           & 22,800 & TTS, VC           
\\ \midrule
\multicolumn{5}{c}{\centering \textbf{Development}} \\ 
\midrule
        19LA \cite{WANG2020101114}            & English        & 2,548           & 22,296 &  TTS, VC 
\\ \midrule
\multicolumn{5}{c}{\centering \textbf{Evaluation}} \\ 
\midrule
        19LA \cite{WANG2020101114}   & English      & 7,355           & 63,882  &    TTS, VC        \\ \midrule
        \multirow{1}{*}{21} \begin{tabular}[c]{@{}c@{}}LA\\DF\end{tabular}  \cite{10155166}  & English      & \begin{tabular}[c]{@{}c@{}}14,816\\14,869\end{tabular}           & \begin{tabular}[c]{@{}c@{}}133,360\\519,059\end{tabular} &  TTS, VC         \\ \midrule

        FoR \cite{8906599}   & English      & 2,264           & 2,370  &     TTS       \\\midrule
        ITW \cite{muller22_interspeech} & English      & 19,963           & 11,816 &   Unknown          \\\midrule
        DFADD \cite{10832250} & English      & 755           & 3,000   &    TTS       \\\midrule
        Librisevoc \cite{Sun_2023_CVPR} & English      & 2,641           & 15,846  & Vocoded           \\\midrule

        \multirow{1}{*}{DECRO} \begin{tabular}[c]{@{}c@{}}EN\\CH\end{tabular} \cite{10.1145/3543507.3583222}& \begin{tabular}[c]{@{}c@{}}English\\Chinese\end{tabular}      & \begin{tabular}[c]{@{}c@{}}4,306\\6,109\end{tabular}           & \begin{tabular}[c]{@{}c@{}}14,884\\12,015\end{tabular} &  TTS, VC         \\ \midrule
        \multirow{1}{*}{ADD23} \begin{tabular}[c]{@{}c@{}}R1\\R2\end{tabular}   \cite{DBLP:conf/dada/YiTFYWWZZZRXZGW23}  & Chinese      & \begin{tabular}[c]{@{}c@{}}80,000\\87,500\end{tabular}
        & \begin{tabular}[c]{@{}c@{}}31,976\\30,977\end{tabular} &  TTS, VC         \\ \midrule
        HABLA \cite{tamayoflorez23_interspeech} & Spanish      & 9,057           & 23,270   &  TTS, VC         \\\midrule
        \multirow{1}{*}{MLAAD} \begin{tabular}[c]{@{}c@{}}EN\\FR\\DE\\ES\\IT\\PL\\UK\\RU\end{tabular} \cite{10650962}   & \begin{tabular}[c]{@{}c@{}}English\\French\\German\\Spanish\\Italian\\Polish\\Ukrainian\\Russian\end{tabular}       & \begin{tabular}[c]{@{}c@{}}28,233\\7,686\\8,696\\6,655\\7,611\\5,489\\4,709\\4,540\end{tabular} 
        & \begin{tabular}[c]{@{}c@{}}36,000\\8,000\\9,000\\7,000\\8,000\\6,000\\5,000\\5,000\end{tabular}  &  TTS         \\
        \bottomrule
    \end{tabular}
    }
    \label{tab:dataset_statistics}
\end{table}

\subsection{Performance Metric}
To evaluate the performance of our model, we employ the commonly used metric Equal Error Rate (EER). The EER corresponds to the point where the False Acceptance (FA) rate (\( P_{\text{fa}}^{\text{CM}} \), false alarm when spoofed trials misclassified as bona fide) and the False Rejection (FR) rate (\( P_{\text{miss}}^{\text{CM}} \), miss when bona fide trials misclassified as spoofed) are equal. These rates are computed as follows:
\begin{equation}
    P_{\text{fa}}^{\text{CM}} (\tau_{\text{CM}}) = \frac{\# \text{ spoofed trials with CM scores } > \tau_{\text{CM}}}{\# \text{spoofed trials}},
\end{equation}
\begin{equation}
    P_{\text{miss}}^{\text{CM}} (\tau_{\text{CM}}) = \frac{\# \text{ bona fide trials with CM scores } \leq \tau_{\text{CM}}}{\#\text{ bona fide trials}}.
\end{equation}
A FA occurs when a spoofed trial receives a classification score greater than the threshold \( \tau_{\text{CM}} \) and is incorrectly accepted as bona fide. Conversely, a FR happens when a bona fide trial receives a score less than or equal to \( \tau_{\text{CM}} \) and is mistakenly rejected. As for the metric value, the lower EER indicates a better performance of the model.

Usually, the detection model outputs two confident scores to indicate the possibility of one audio being bona fide or spoofed. The LogLikelihood Ratio (LLR) will be saved as the final score of this audio, formulated as:
\begin{equation}
    \text{LLR}_t = \log p(X_t | \mathcal{H}_0) - \log p(X_t | \mathcal{H}_1),
\end{equation}

where \( X_t \) represents the audio segment corresponding to the \( t \)-th trial. The hypotheses are defined as follows: \( \mathcal{H}_0 \) denotes the null hypothesis, indicating that \( X_t \) is a bona fide speech segment, while \( \mathcal{H}_1 \) represents the alternative hypothesis, implying that \( X_t \) is a spoofed speech segment.

\subsection{Experimental Setup}

We utilize the pretrained XLS-R model from Huggingface\footnote{\url{https://huggingface.co/facebook/wav2vec2-xls-r-300m}}. Audio inputs are dynamically padded to match the length of the longest sample within a batch of size 5. During training, we set the learning rate to \( 3 \times 10^{-6} \) and employ the Adam optimizer with a weight decay of \( 1 \times 10^{-4} \). To address the class imbalance in the dataset, we apply a weighted  $\mathcal{L}_{\text{CE}}$ loss, assigning a weight of 0.9 to the minority class (bona fide) and 0.1 to the majority class (spoofed). Models are fine-tuned with a patience of 3 epochs, and the model that performs best on the 19LA development set is selected for evaluation. We set the embedding of the feature projection to 128 and employed 4 MultiConv layers, inspired by \cite{rosello23_interspeech, truong24b_interspeech}. To assess our models on multiple evaluation datasets, we use a batch size of 1 to evaluate the full utterance without padding. All trainings and evaluations were conducted on a single A100 GPU. 
We also incorporate data augmentation techniques, as outlined in RawBoost \cite{9746213}, to enhance the model's robustness. These techniques include linear and nonlinear convolutive noise, impulsive signal-dependent additive noise, stationary signal-independent additive noise, and randomly colored noise. To validate our approach, we did experiments 
 by selecting different MultiConv configurations. Next, we optimized the training objective by adding \(\mathcal{L}_{\text{CKA}}\) in combination with \(\mathcal{L}_{\text{CE}}\). Finally, we performed an ablation study to assess the contribution of each component. 
 
\subsection{Comparison with State-Of-The-Art}
\begin{table}[]
\centering
\caption{Overall performance comparison to SOTA systems across multiple datasets such as 19LA, 21LA, 21DF, and ITW evaluation sets. Bold font indicates best results. (*) denotes our average reproduced results obtained from three runs.}
\label{tab:sota_comparison}
\resizebox{\columnwidth}{!}{%
\begin{tabular}{lccccc}
\toprule
\multirow{2}{*}{\textbf{Systems}} & \textbf{19LA} & \textbf{21LA} & \textbf{21DF} & \textbf{ITW} & \multirow{2}{*}{\begin{tabular}[c]{@{}c@{}}\textbf{Params}\\ (M)\end{tabular}} \\ 
\cline{2-5}
& EER\,$\downarrow$ & EER\,$\downarrow$ & EER\,$\downarrow$ & EER\,$\downarrow$ & \\
\midrule
WavLM+MFA \cite{10447923} & 0.42 & 5.08 & 2.56 & -- & -- \\     
WavLM+AttM \cite{pan24c_interspeech} & 0.65 & 3.50 & 3.19 & -- & -- \\
XLS-R+MoE \cite{wang2024mixture} & 0.74 & 2.96 & 2.54 & 12.48 & 341 \\
XLS-R+AASIST \cite{tak22_odyssey} & -- & 0.82 & 2.85 & -- & -- \\
XLS-R+AASIST2 \cite{10448049} & 0.15 & 1.61 & 2.77 & -- & -- \\
XLS-R+Conformer+TCM \cite{truong24b_interspeech} & -- & 1.03 & 2.06 & -- & 319 \\
XLS-R+SLS \cite{zhang2024audio} & -- & 2.87 & 1.92 & 7.46 & -- \\
XLS-R+LSR+LSA \cite{10888328} & 0.12 & 1.05 & 1.86 & 5.54 & -- \\
XLS-R+DuaBiMamba \cite{10909468} & -- & 0.93 & 1.88 & 6.71 & 319 \\
XLS-R+WavSpec \cite{10890563} & -- & -- & 1.90 & 6.58 & -- \\
XLS-R+STCA+LMDC \cite{10889337} & 0.09 & \textbf{0.78} & 1.87 & -- & -- \\
\midrule
XLS-R+MultiConv (Proposed) & \textbf{0.08}\,(0.10)$^{*}$ & 2.77\,(2.76)$^{*}$ & \textbf{1.43}\,(1.53)$^{*}$ & \textbf{4.44}\,(4.78)$^{*}$ & 318 \\
\bottomrule
\end{tabular}%
}
\end{table}

\begin{table*}[!t]
\centering
\caption{Overall performance comparison with SOTA systems across datasets grouped by language family. Bold font indicates best results. \(\dagger\) are results evaluated using released checkpoint.}
\label{tab:sota_by_family}
\resizebox{\textwidth}{!}{%
\begin{tabular}{l c c c c c c c c c c c c c c c c c c}
\toprule
\multirow{3}{*}{\textbf{Systems}} & 
\multicolumn{7}{c}{\textbf{Germanic}} & 
\multicolumn{4}{c}{\textbf{Romance}} & 
\multicolumn{3}{c}{\textbf{Slavic}} &
\multicolumn{3}{c}{\textbf{Sino-Tibetan}} & 
\multirow{3}{*}{\begin{tabular}[c]{@{}c@{}}\textbf{Params}\\ (M)\end{tabular}} \\

\cmidrule(lr){2-8} 
\cmidrule(lr){9-12}
\cmidrule(lr){13-15}
\cmidrule(lr){16-18}

& \textbf{ITW} & \textbf{FoR} & \textbf{D-EN} & \textbf{DFADD} & \textbf{Librisevoc} & \textbf{M-EN} & \textbf{M-DE} 
& \textbf{M-FR} & \textbf{M-IT} & \textbf{M-ES} & \textbf{HABLA} 
& \textbf{M-PL} & \textbf{M-UK} & \textbf{M-RU} & \textbf{ADD23-R1} & \textbf{ADD23-R2} & \textbf{D-CH}
& \\
  \cline{2-18}
  & EER$\downarrow$ & EER$\downarrow$ & EER$\downarrow$ & EER$\downarrow$ & EER$\downarrow$ & EER$\downarrow$ & EER$\downarrow$
  & EER$\downarrow$ & EER$\downarrow$ & EER$\downarrow$ & EER$\downarrow$
  & EER$\downarrow$ & EER$\downarrow$ & EER$\downarrow$
  & EER$\downarrow$ & EER$\downarrow$ & EER$\downarrow$
  &     \\ 
\midrule
Conformer+TCM \cite{truong24b_interspeech} \(\dagger\) & 7.79 & 12.15 & 1.77 & 8.87 & 2.35 & 14.35 & 20.59 & 7.06 & 7.45 & 11.75 & 2.28 & 11.86 & 21.75 & 7.62 & 23.42 & 22.74 & 12.88 & 319 \\
SLS \cite{zhang2024audio} \(\dagger\) & 7.46 & 6.71 & \textbf{1.86} & 7.53 & 1.97 & 15.59 & 19.71 & 6.44 & 6.73 & 9.69 & 1.62 & 8.67 & 21.13 & 8.64 & \textbf{19.37} & 21.09 & \textbf{12.26} & 340 \\
DuaBiMamba \cite{10909468} \(\dagger\) &  6.71 & \textbf{1.51} & 4.53 & 15.87 & 6.78 & \textbf{9.52} & 23.57 & 11.23 & 9.17 & 15.83 & 8.06 & 16.09 & 11.89 & 14.05 & 27.59 & 28.69 & 17.32 & 319\\

\midrule

MultiConv (Proposed) & \textbf{4.44} & 5.66 & 2.26 & \textbf{6.60} & \textbf{1.70} & 13.56 & \textbf{14.37} & \textbf{6.24} & \textbf{4.77} & \textbf{6.97} & \textbf{1.45} & \textbf{8.53} & \textbf{10.18} & \textbf{4.76} & 20.28 & \textbf{17.58} & 13.68 & 318 \\

\bottomrule
\end{tabular}%
}
\end{table*}

As shown in Table \ref{tab:sota_comparison}, our proposed XLS-R+MultiConv model achieves SOTA performance on multiple in-domain datasets, including 19LA (0.08\%) and 21DF (1.43\%), demonstrating the effectiveness of the proposed method compared to other SOTA systems. However, it performs worse on 21LA (2.77\%). For out-of-domain data, our system also sets a new benchmark, achieving a 4.44\% EER, which indicates improved robustness to real-world conditions while maintaining a compact model size of only 318M parameters. Contrary to most SOTA models that are trained and evaluated using 4-second audio segments (64,600 samples), our model, which is trained on full utterances, achieves lower EER in most datasets. This demonstrates that using MultiConv as a back-end classifier enables the learning of fine-grained local and global discriminative features.

\begin{figure}[tb!]
    \centering
    \includegraphics[width=\columnwidth]{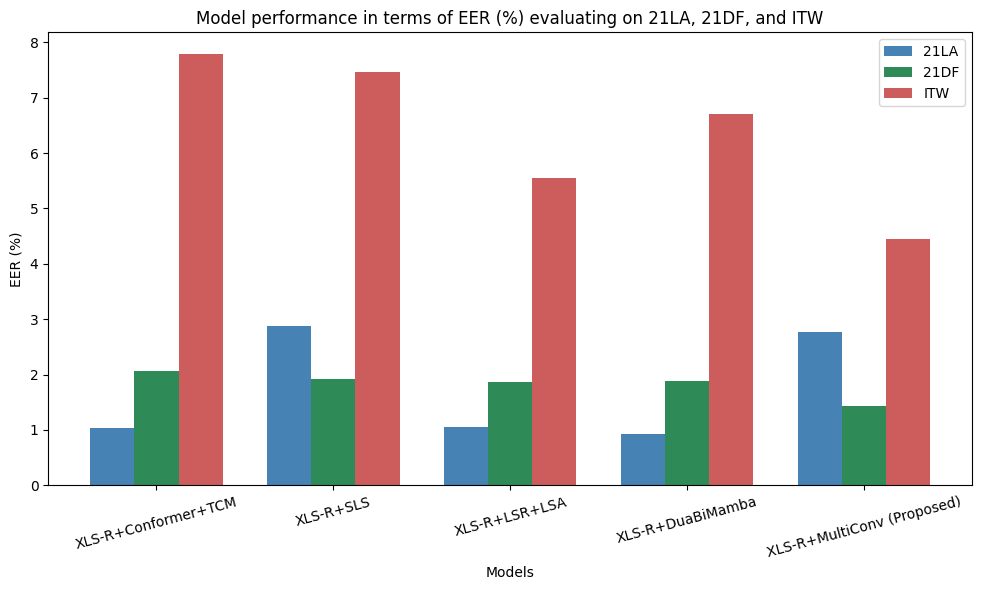}
    \caption{Top 5 models' performance in terms of EER (\%) on 21LA (blue), 21DF (green), and ITW (red) datasets.}
    \label{fig:model_performance}
\end{figure}

Figure \ref{fig:model_performance} shows the performance of the top 5 best models on the 21LA, 21DF, and ITW datasets. The XLS-R+Conformer+TCM, XLS-R+DuaBiMamba, and XLS-R+LSR+LSA models use the last XLS-R output's contextual Transformer layer, with the first employing a Conformer-based architecture, the second using a Mamba-based architecture, and the last incorporating a graph-based AASIST approach. The first two models are based on an averaged checkpoint of the top 5 validation models, while the third model uses codec augmentation as an additional data augmentation technique. For the SLS model, the 24 contextual Transformer layers from XLS-R are selectively employed.

We observe that leveraging all Transformer layers from XLS-R achieves better performance compared to the last output following a Conformer-based classifier on the 21DF and ITW datasets. Since the Conformer-based, graph-based, and Mamba classifiers were trained for specific tasks, they performed well on the 21LA, and 21DF task individually. The Mamba-based classifier demonstrated promising results for both the 21DF and ITW datasets. However, our model outperformed all these models on both in-domain and out-of-domain datasets while having a competitive performance on 21LA with XLS-R+SLS. This demonstrates that by using a single model and fully exploiting the capabilities of XLS-R's transformer layers with the MultiConv classifier and CKA loss, our proposed system specializes and generalizes well, achieving low EER.

Table~\ref{tab:sota_by_family} presents a comprehensive comparison of SOTA systems with released checkpoints evaluated on datasets grouped by language family such as Germanic, Romance, Slavic, and Sino-Tibetan. Among the evaluated systems, the proposed XLS-R+MultiConv model consistently achieves strong performance across all branches in most datasets. Within the Germanic group, XLS-R+MultiConv outperforms all baselines on ITW (4.44\%), DFADD (6.60\%), Librisevoc (1.70\%), and M-DE (14.37\%), while achieving competitive results on FoR (5.66\%) and M-EN (13.56\%). In the Romance group, it leads on all datasets, including M-FR (6.24\%), IT (4.77\%), and ES (6.97\%), showing its robustness across Latin-based languages. In the Slavic branch, XLS-R+MultiConv again demonstrates superior performance on M-RU (4.76\%) and performs competitively on M-PL (8.53\%) and UK (10.18\%). For Sinothe Sino-Tibetanguage family, particular Chinese datasets, all models remain approximately the same performance on ADD23-R1, R2, and D-CH evaluation sets. Compared to the strongest alternative, XLS-R+SLS, the proposed method achieves lower EERs in 14 out of 17 out-of-domain evaluation sets, while also maintaining a smaller back-end parameter.

\section{Result Analysis and Ablation Study}
In this section, we first analyze the impact of different MultiConv kernel configurations. Next, we evaluate the effectiveness of CKA as a loss function. We then examine the in-domain dataset to better understand why our system struggles to detect certain attacks. For the out-of-domain analysis, we group the evaluation by language. Finally, we conduct an ablation study to further investigate the contributions of each component.

\subsection{Impact of Kernel Configurations}
Our analysis of MultiConv architectures with varying kernel configurations reveals critical trade-offs between kernel size and performance across datasets. While smaller kernels \{3, 7\} excel in capturing artifacts, achieving very low EER on ITW (4.04\% EER) and Librisevoc (1.36\% EER), they underperform on other task such as 21LA (4.38\% EER). Conversely, larger kernels \{19, 23\} demonstrate superior performance on datasets like 19LA (0.09\% EER) and HABLA (1.77\% EER). More multi-kernel configurations \{3, 7, 11, 15\} achieve the best generalization on Chinese challenges ADD23 in both scenarios R1 and R2 (21.75\% and 21.98\% EER respectively). Notably, no single configuration universally outperforms others and task-specific kernel selection is needed. Combinations of larger kernels \{19, 23, 27, 31\} excel on FoR (2.44\% EER) but underperform on other datasets. These findings underscore the importance of hierarchical receptive fields in SSD, where adaptive kernel ensembles mitigate domain shifts and enhance robustness.

\subsection{Efficiency of CKA}
Our investigation comparing the use of $\mathcal{L}_{\text{CE}}$ with and without $\mathcal{L}_{\text{CKA}}$ reveals that incorporating CKA significantly enhances robustness and cross-domain generalization in SDD. While $\mathcal{L}_{\text{CE}}$ alone achieves strong performance on simpler, clean datasets such as 19LA (0.09\% EER using \{19,23\}-kernels), the combined objective $\mathcal{L}_{\text{CE}} + \mathcal{L}_{\text{CKA}}$ generalizes better on more complex in-domain datasets like 21LA and 21DF, achieving 2.55\% and 1.75\% EER, respectively. This indicates that the joint loss encourages the model to learn more diverse and complementary features across layers, reducing redundancy and enhancing robustness to within-domain complexity. On out-of-domain datasets, the addition of $\mathcal{L}_{\text{CKA}}$ also consistently improves performance, particularly for English-language data. For datasets with different linguistic characteristics, the model with \{3, 7, 11, 15\}-kernels performs well on HABLA (1.45\% EER), ADD23-R2 (17.58\%), and remains mixed results on D-CH and ADD23-R1.

\subsection{In-domain Analysis}
\begin{figure}[tb!]
    \centering
    \includegraphics[width=1.075\columnwidth]{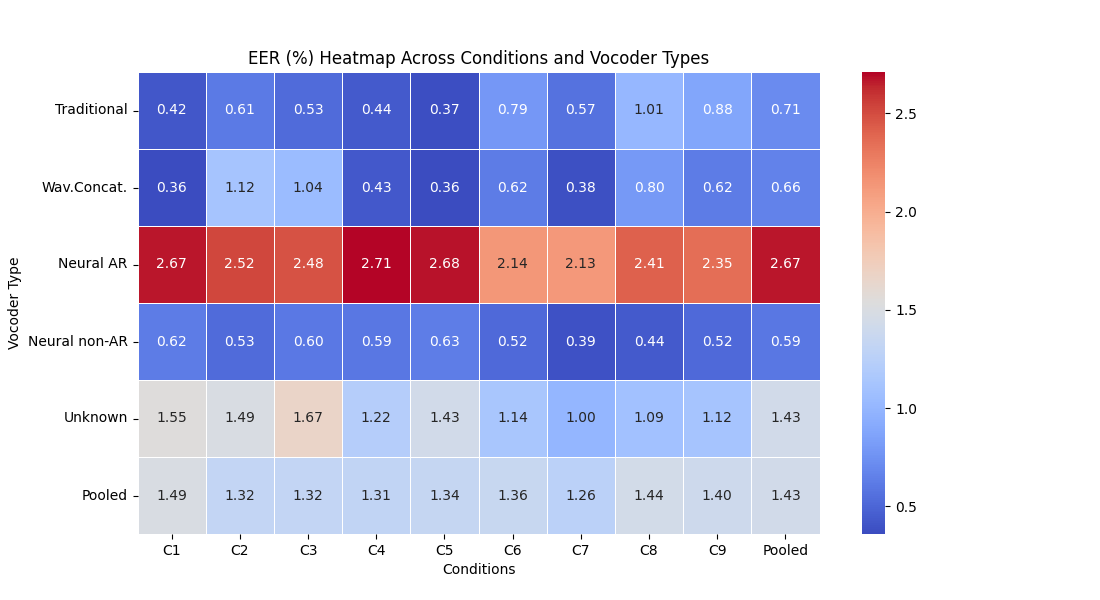}
    \caption{Heatmap of performance (EER \%) of our system with evaluated on 21DF evaluation set. “Wav.Concat.” denotes waveform concatenation and AR
denotes autoregressive.}
    \label{fig:df_analysis}
\end{figure}

\begin{figure}[tb!]
    \centering
    \includegraphics[width=1.0825\columnwidth]{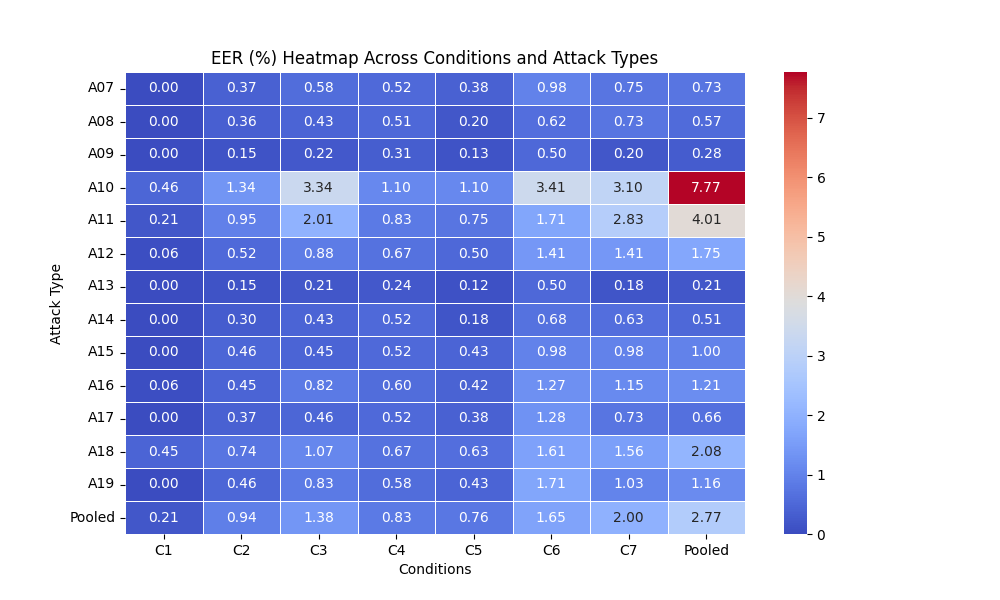}
    \caption{Heatmap of performance (EER \%) of our system with evaluated on 21LA evaluation set. A07 to A16 denotes TTS–based attacks, and A17 to
A19 denotes VC–based attacks.}
    \label{fig:la_analysis}
\end{figure}

For in-domain performance analysis, Figure \ref{fig:df_analysis} shows the EER performance across different conditions (C1-C9) using various types of vocoders, as described in \cite{10155166}. The results indicate that the model struggles when faced with neural AutoRegressive (AR) vocoders, highlighting the need for further refinement to address this type of attack. However, our model achieves very low EER for most other vocoder types, including unknown vocoders, demonstrating its generalizability to unseen attacks with 1.43\% of the pooled EER.
Furthermore, Figure \ref{fig:la_analysis} presents our model's performance on 21LA with different attacks (A07-A19) across various conditions (C1-C9) \cite{10155166}. For TTS-based attacks, the model struggles particularly with A10, which uses the neural vocoder WaveRNN in combination with Tacotron 2 with the highest EER (7.77\%). In contrast, A11, another neural TTS system similar to A10 but employing the Griffin-Lim algorithm for waveform generation, is less difficult (4.01\% EER). For VC-based attacks, our model performs well, achieving low EER across A17-A19, demonstrating that even when the linguistic content is identical, our model can effectively distinguish between spoofed and bona fide samples.

\begin{table*}[!t]
\centering
\caption{Ablation study of MultiConv kernel sizes and different components in our proposed method. Dark cells indicate the same model. Bold font indicates best results.}
\label{tab:sota}
\resizebox{\textwidth}{!}{%
\begin{tabular}{lccccccccccccc}
\toprule
  \multirow{2}{*}{\textbf{MultiConv kernel sizes}} &
  \multirow{1}{*}{\textbf{19LA}} &
  \multirow{1}{*}{\textbf{21LA}} &
  \multirow{1}{*}{\textbf{21DF}} &
  \multirow{1}{*}{\textbf{FoR}} &
  \multirow{1}{*}{\textbf{ITW}} &
  \multirow{1}{*}{\textbf{DFADD}} &
  \multirow{1}{*}{\textbf{Librisevoc}} &
  \multirow{1}{*}{\textbf{D-EN}} &
  \multirow{1}{*}{\textbf{D-CH}} &
  \multirow{1}{*}{\textbf{ADD23-R1}} &
  \multirow{1}{*}{\textbf{ADD23-R2}} &
  \multirow{1}{*}{\textbf{HABLA}} &
   \\
    \cline{2-14}
  & EER$\downarrow$  & EER$\downarrow$  & EER$\downarrow$  & EER$\downarrow$  & EER$\downarrow$  & EER$\downarrow$   & EER$\downarrow$   & EER$\downarrow$   & EER$\downarrow$   & EER$\downarrow$   & EER$\downarrow$   & EER$\downarrow$   &     \\ \midrule
\multicolumn{14}{c}{\centering \textbf{$\mathcal{L}_{\text{CE}}$}} \\ 
\midrule
\{3, 7\} & 0.15 & 4.38 & 2.12 & 7.64 & \textbf{4.04} & 7.41 & \textbf{1.36} & 1.19 & 15.36 & 24.58 & 27.41 & 2.02 \\

\{11, 15\} & 0.95 & 3.11 & 2.22 & 4.14 & 5.85 & 13.77 & 3.30 & 2.90 & 15.03 & 25.06 & 28.97 & 2.18 \\

\{19, 23\} & \textbf{0.09} & 3.61 & 2.08 & 3.17 & 6.51 & \textbf{7.28} & 1.67 & 1.83 & 9.83 & 23.08 & 23.99 & \textbf{1.77} \\

\{27, 31\} & 0.27 & \textbf{2.02} & 2.33 & 8.18 & 5.29 & 9.29 & 2.35 & 2.72 & 14.24 & 22.35 & 23.50 & 2.05 \\

\{3, 7, 11, 15\} & 0.91 \cellcolor{gray!75}& 3.61 \cellcolor{gray!75}& \textbf{2.02} \cellcolor{gray!75}& 2.53 \cellcolor{gray!75}& 5.51 \cellcolor{gray!75}& 23.69 \cellcolor{gray!75}& 4.02 \cellcolor{gray!75}& 1.93 \cellcolor{gray!75}& 13.87 \cellcolor{gray!75}& \textbf{21.75} \cellcolor{gray!75}& \textbf{21.98} \cellcolor{gray!75}& 1.91 \cellcolor{gray!75}\\

\{11, 15, 19, 23\} & 0.30 & 3.94 & 2.80 & 6.54 & 5.15 & 13.01 & 2.20 & \textbf{1.02} & \textbf{6.96} & 22.33 & 23.62 & 2.09 \\

\{19, 23, 27, 31\} & 2.61 & 4.55 & 2.39 & \textbf{2.44} & 5.61 & 25.57 & 1.98 & 4.85 & 14.23 & 24.66 & 26.19 & 2.94  \\

\midrule
\multicolumn{14}{c}{\centering \textbf{$\mathcal{L}_{\text{CE}} + \mathcal{L}_{\text{CKA}}$}} \\ 
\midrule

\{3, 7\} & 0.22 & 4.23 & 1.79 & \textbf{2.20} & 4.53 & 9.25 & 1.59 & 1.81 & 13.59 & 23.57 & 24.00 & 2.25 \\

\{11, 15\} & 0.18 & 3.24 & 1.56 & 3.67 & 4.44 & \textbf{5.70} & 1.63 & \textbf{1.04} & 13.13 & 22.81 & 20.58 & 1.88 \\

\{19, 23\} & 0.20 & \textbf{2.55} & 1.75 & 2.87 & 5.20 & 8.64 & 1.40 & 1.72 & 14.21 & 23.35 & 24.15 & 1.66 \\

\{27, 31\} & 0.18 & 3.38 & 1.86 & 4.86 & 4.73 & 8.74 & 1.63 & 1.35 & 15.27 & 23.50 & 25.39 & 2.14 \\

\{3, 7, 11, 15\} & \textbf{0.08} \cellcolor{gray!25}& \textbf{2.77} \cellcolor{gray!25}& \textbf{1.43} \cellcolor{gray!25}& 5.66 \cellcolor{gray!25}& 4.44 \cellcolor{gray!25}& 6.60 \cellcolor{gray!25}& 1.70 \cellcolor{gray!25}& 2.26 \cellcolor{gray!25}& 13.68 \cellcolor{gray!25}& \textbf{20.28} \cellcolor{gray!25}& \textbf{17.58} \cellcolor{gray!25}& \textbf{1.45} \cellcolor{gray!25}\\

\{11, 15, 19, 23\} & 0.18 & 3.05 & 1.93 & 8.35 & 4.90 & 15.24 & 1.93 & 2.79 & 13.87 & 23.77 & 25.85 & 1.93 \\

\{19, 23, 27, 31\} & 0.16 & 2.69 & 1.77 & 2.96 & \textbf{4.39} & 10.73 & \textbf{1.02} & 1.12 & \textbf{12.15} & 20.78 & 21.93 & 1.60 \\
\midrule
\multicolumn{14}{c}{\centering \textbf{Ablation Study}} \\ 

\midrule
\multirow{3}{*}{ \begin{tabular}[c]{@{}c@{}}3 runs\\\{3, 7, 11, 15\}\end{tabular}} & 0.09  & 3.14  & 1.48 & \textbf{1.81} & 4.94 & \textbf{5.68} & 1.77  & 1.23  & 13.85 & 22.48 & 19.51  & \textbf{1.25} \\
 & \textbf{0.08} \cellcolor{gray!25}& 2.77 \cellcolor{gray!25}& \textbf{1.43} \cellcolor{gray!25}& 5.66 \cellcolor{gray!25}& \textbf{4.44} \cellcolor{gray!25}& 6.60 \cellcolor{gray!25}& 1.70 \cellcolor{gray!25}& 2.26 \cellcolor{gray!25}& 13.68 \cellcolor{gray!25}& 20.28 \cellcolor{gray!25}& \textbf{17.58} \cellcolor{gray!25}& 1.45 \cellcolor{gray!25}\\
  & 0.12 & 2.38 & 1.68 & 3.80 & 4.97 & 7.96 & 1.89 & 2.55 & \textbf{11.44} & \textbf{19.26} & 18.23 & 1.54 \\
\midrule
\quad w/o $\mathcal{L}_{\text{CKA}}$ & 0.91 \cellcolor{gray!75} & 3.61 \cellcolor{gray!75}& 2.02 \cellcolor{gray!75}& 2.53 \cellcolor{gray!75}& 5.51 \cellcolor{gray!75}& 23.69 \cellcolor{gray!75}& 4.02 \cellcolor{gray!75}& 1.93 \cellcolor{gray!75}& 13.87 \cellcolor{gray!75}& 21.75 \cellcolor{gray!75}& 21.98 \cellcolor{gray!75}& 1.91 \cellcolor{gray!75}\\
\quad w/o SwiGLU  & 0.12 & 3.56 & 1.94 & 4.51 & 5.26 & 7.13 & \textbf{1.24} & 2.01 & 14.55 & 26.08 & 26.32 & 2.13 \\
\quad w/o DA & 0.18 & 8.48 & 3.71 & 6.28 & 5.47 & 11.68 & 1.40 & 1.44 & 17.50 & 21.73 & 26.20 & 2.67 \\
\midrule
\quad 8 layers & 0.09 & 2.36 & 2.05 & 1.81 & 5.06 & 11.79 & 2.04 & 3.32 & 12.97 & 26.70 & 28.67 & 1.69 & \\
\quad 12 layers & 0.14 & \textbf{1.61} & 2.87 & 4.42 & 5.53 & 14.30 & 2.07 & \textbf{1.12} & 15.40 & 28.54 & 30.98 & 1.80 & \\
\bottomrule
\end{tabular}%
}
\end{table*}

\subsection{Out-of-domain Analysis}
Despite being trained exclusively on English-language data, the proposed model demonstrates notable generalization capabilities across a diverse set of languages, as evidenced by its EER on out-of-domain datasets. Performance remains good on Germanic languages (e.g., DECRO EN: 2.26\%, Librisevoc: 1.70\%), suggesting robustness to phonetic and acoustic variations within closely related Indo-European language groups. However, a performance drop is observed when the model is confronted with recent TTS-based attacks from M-EN and diffusion-based attacks from DFADD and the diversity in the dataset from FoR. Despite their phonetic similarity to English, German samples (EER: 14.37\%) remain particularly challenging. The model also struggles with languages more distant from English, especially in the Slavic group (e.g., M-PL: 8.53\%, M-UK: 10.18\%), where increased EER suggests difficulties in transferring learned representations for effective spoof detection. Nonetheless, relatively low EER on Romance-language datasets (e.g., HABLA: 1.45\%, M-IT: 4.77\%) indicate a degree of cross-family generalization, suggesting that the model captures some language-independent spoofing cues. The difference in EER between HABLA and M-ES may be attributable to differences in attack techniques. For Sino-Tibetan languages, particularly Chinese, the model consistently underperforms across all three evaluated datasets (D-CH, ADD23-R1 and R2), further highlighting the limitations of monolingual training when facing typologically distant languages. These results underscore both the potential and the limitations of monolingual training for building generalized spoofing detection systems.

We conducted an ablation study to assess the contribution of each component in our proposed architecture. Across three independent runs, the model consistently achieved stable and comparable results. Removing the SwiGLU activation function led to a degradation in performance, although the results remained competitive. Notably, the exclusion of data augmentation resulted in a substantial drop in performance, highlighting the critical role of diverse and complex training data in promoting generalization. We also experimented with increasing the number of MultiConv layers. While deeper models exhibited improved performance on 21LA (e.g., 2.36\% EER with 8 layers and 1.61\% EER with 12 layers), they failed to generalize effectively across other datasets.

\section{Conclusion}
In this study, we proposed a novel approach to audio deepfake detection by leveraging the full potential of XLS-R hidden representations through a gating mechanism, combined with gated MultiConv layers as a back-end classifier. We demonstrated that using Centered Kernel Alignment as a loss function encourages inter-layer dissimilarity, enabling the learning of diverse and complementary representations. This strategy significantly improves the model’s robustness across both in-domain and out-of-domain datasets, spanning multiple language families. Our results further emphasize the critical role of training data diversity both in acoustic conditions and linguistic content for achieving generalization in real-world scenarios. Models trained exclusively on clean data exhibit limited performance when confronted with realistic, heterogeneous deepfake attacks. Future work will explore multilingual and noisy training data to further improve cross-domain generalization and detection accuracy.

\begin{acks}
This work was granted access to the HPC/AI resources of
IDRIS under the allocation 2023–AD011013889R1 made by
GENCI and funded by Côtes d’Armor departmental council and
by ANR (Agence Nationale de la Recherche) through Doctoral
Contract as part of the ANR–20–THIA–0018 project.
\end{acks}

\bibliographystyle{ACM-Reference-Format}
\balance
\bibliography{ref}


\begin{thebibliography}{66}


\ifx \showCODEN    \undefined \def \showCODEN     #1{\unskip}     \fi
\ifx \showISBNx    \undefined \def \showISBNx     #1{\unskip}     \fi
\ifx \showISBNxiii \undefined \def \showISBNxiii  #1{\unskip}     \fi
\ifx \showISSN     \undefined \def \showISSN      #1{\unskip}     \fi
\ifx \showLCCN     \undefined \def \showLCCN      #1{\unskip}     \fi
\ifx \shownote     \undefined \def \shownote      #1{#1}          \fi
\ifx \showarticletitle \undefined \def \showarticletitle #1{#1}   \fi
\ifx \showURL      \undefined \def \showURL       {\relax}        \fi
\providecommand\bibfield[2]{#2}
\providecommand\bibinfo[2]{#2}
\providecommand\natexlab[1]{#1}
\providecommand\showeprint[2][]{arXiv:#2}

\bibitem[Abdelmajid H.~Mansour(2015)]%
        {10.5120/20312-2362}
\bibfield{author}{\bibinfo{person}{Khalid A.~Mohammed Abdelmajid H.~Mansour,
  Gafar Zen Alabdeen~Salh}.} \bibinfo{year}{2015}\natexlab{}.
\newblock \showarticletitle{Voice Recognition using Dynamic Time Warping and
  Mel-Frequency Cepstral Coefficients Algorithms}.
\newblock \bibinfo{journal}{\emph{International Journal of Computer
  Applications}} \bibinfo{volume}{116}, \bibinfo{number}{2}
  (\bibinfo{date}{April} \bibinfo{year}{2015}), \bibinfo{pages}{34--41}.
\newblock
\showISSN{0975-8887}
\href{https://doi.org/10.5120/20312-2362}{doi:\nolinkurl{10.5120/20312-2362}}


\bibitem[Ardila et~al\mbox{.}(2020)]%
        {ardila-etal-2020-common}
\bibfield{author}{\bibinfo{person}{Rosana Ardila}, \bibinfo{person}{Megan
  Branson}, \bibinfo{person}{Kelly Davis}, \bibinfo{person}{Michael Kohler},
  \bibinfo{person}{Josh Meyer}, \bibinfo{person}{Michael Henretty},
  \bibinfo{person}{Reuben Morais}, \bibinfo{person}{Lindsay Saunders},
  \bibinfo{person}{Francis Tyers}, {and} \bibinfo{person}{Gregor Weber}.}
  \bibinfo{year}{2020}\natexlab{}.
\newblock \showarticletitle{Common Voice: A Massively-Multilingual Speech
  Corpus}. In \bibinfo{booktitle}{\emph{Proceedings of the Twelfth Language
  Resources and Evaluation Conference}},
  \bibfield{editor}{\bibinfo{person}{Nicoletta Calzolari},
  \bibinfo{person}{Fr{\'e}d{\'e}ric B{\'e}chet}, \bibinfo{person}{Philippe
  Blache}, \bibinfo{person}{Khalid Choukri}, \bibinfo{person}{Christopher
  Cieri}, \bibinfo{person}{Thierry Declerck}, \bibinfo{person}{Sara Goggi},
  \bibinfo{person}{Hitoshi Isahara}, \bibinfo{person}{Bente Maegaard},
  \bibinfo{person}{Joseph Mariani}, \bibinfo{person}{H{\'e}l{\`e}ne Mazo},
  \bibinfo{person}{Asuncion Moreno}, \bibinfo{person}{Jan Odijk}, {and}
  \bibinfo{person}{Stelios Piperidis}} (Eds.). \bibinfo{publisher}{European
  Language Resources Association}, \bibinfo{address}{Marseille, France},
  \bibinfo{pages}{4218--4222}.
\newblock
\showISBNx{979-10-95546-34-4}
\urldef\tempurl%
\url{https://aclanthology.org/2020.lrec-1.520/}
\showURL{%
\tempurl}


\bibitem[Ba et~al\mbox{.}(2023)]%
        {10.1145/3543507.3583222}
\bibfield{author}{\bibinfo{person}{Zhongjie Ba}, \bibinfo{person}{Qing Wen},
  \bibinfo{person}{Peng Cheng}, \bibinfo{person}{Yuwei Wang},
  \bibinfo{person}{Feng Lin}, \bibinfo{person}{Li Lu}, {and}
  \bibinfo{person}{Zhenguang Liu}.} \bibinfo{year}{2023}\natexlab{}.
\newblock \showarticletitle{Transferring Audio Deepfake Detection Capability
  across Languages}. In \bibinfo{booktitle}{\emph{Proceedings of the ACM Web
  Conference 2023}} (Austin, TX, USA) \emph{(\bibinfo{series}{WWW '23})}.
  \bibinfo{publisher}{Association for Computing Machinery},
  \bibinfo{address}{New York, NY, USA}, \bibinfo{pages}{2033–2044}.
\newblock
\showISBNx{9781450394161}
\href{https://doi.org/10.1145/3543507.3583222}{doi:\nolinkurl{10.1145/3543507.3583222}}


\bibitem[Babu et~al\mbox{.}(2022)]%
        {babu22_interspeech}
\bibfield{author}{\bibinfo{person}{Arun Babu}, \bibinfo{person}{Changhan Wang},
  \bibinfo{person}{Andros Tjandra}, \bibinfo{person}{Kushal Lakhotia},
  \bibinfo{person}{Qiantong Xu}, \bibinfo{person}{Naman Goyal},
  \bibinfo{person}{Kritika Singh}, \bibinfo{person}{Patrick {von Platen}},
  \bibinfo{person}{Yatharth Saraf}, \bibinfo{person}{Juan Pino},
  \bibinfo{person}{Alexei Baevski}, \bibinfo{person}{Alexis Conneau}, {and}
  \bibinfo{person}{Michael Auli}.} \bibinfo{year}{2022}\natexlab{}.
\newblock \showarticletitle{XLS-R: Self-supervised Cross-lingual Speech
  Representation Learning at Scale}. In
  \bibinfo{booktitle}{\emph{Interspeech}}.
\newblock
\showISSN{2958-1796}
\href{https://doi.org/10.21437/Interspeech.2022-143}{doi:\nolinkurl{10.21437/Interspeech.2022-143}}


\bibitem[Baevski et~al\mbox{.}(2020)]%
        {NEURIPS2020_92d1e1eb}
\bibfield{author}{\bibinfo{person}{Alexei Baevski}, \bibinfo{person}{Yuhao
  Zhou}, \bibinfo{person}{Abdelrahman Mohamed}, {and} \bibinfo{person}{Michael
  Auli}.} \bibinfo{year}{2020}\natexlab{}.
\newblock \showarticletitle{wav2vec 2.0: A Framework for Self-Supervised
  Learning of Speech Representations}. In \bibinfo{booktitle}{\emph{Advances in
  NeurIPS}}, Vol.~\bibinfo{volume}{33}.
\newblock


\bibitem[Chen et~al\mbox{.}(2022)]%
        {9814838}
\bibfield{author}{\bibinfo{person}{Sanyuan Chen}, \bibinfo{person}{Chengyi
  Wang}, \bibinfo{person}{Zhengyang Chen}, \bibinfo{person}{Yu Wu},
  \bibinfo{person}{Shujie Liu}, \bibinfo{person}{Zhuo Chen},
  \bibinfo{person}{Jinyu Li}, \bibinfo{person}{Naoyuki Kanda},
  \bibinfo{person}{Takuya Yoshioka}, \bibinfo{person}{Xiong Xiao},
  \bibinfo{person}{Jian Wu}, \bibinfo{person}{Long Zhou}, \bibinfo{person}{Shuo
  Ren}, \bibinfo{person}{Yanmin Qian}, \bibinfo{person}{Yao Qian},
  \bibinfo{person}{Jian Wu}, \bibinfo{person}{Michael Zeng},
  \bibinfo{person}{Xiangzhan Yu}, {and} \bibinfo{person}{Furu Wei}.}
  \bibinfo{year}{2022}\natexlab{}.
\newblock \showarticletitle{WavLM: Large-Scale Self-Supervised Pre-Training for
  Full Stack Speech Processing}.
\newblock \bibinfo{journal}{\emph{IEEE J. STSP}} \bibinfo{volume}{16},
  \bibinfo{number}{6} (\bibinfo{year}{2022}).
\newblock
\href{https://doi.org/10.1109/JSTSP.2022.3188113}{doi:\nolinkurl{10.1109/JSTSP.2022.3188113}}


\bibitem[Chetia~Phukan et~al\mbox{.}(2024)]%
        {chetia-phukan-etal-2024-heterogeneity}
\bibfield{author}{\bibinfo{person}{Orchid Chetia~Phukan},
  \bibinfo{person}{Gautam Kashyap}, \bibinfo{person}{Arun~Balaji Buduru}, {and}
  \bibinfo{person}{Rajesh Sharma}.} \bibinfo{year}{2024}\natexlab{}.
\newblock \showarticletitle{Heterogeneity over Homogeneity: Investigating
  Multilingual Speech Pre-Trained Models for Detecting Audio Deepfake}. In
  \bibinfo{booktitle}{\emph{Findings of the Association for Computational
  Linguistics: NAACL 2024}}, \bibfield{editor}{\bibinfo{person}{Kevin Duh},
  \bibinfo{person}{Helena Gomez}, {and} \bibinfo{person}{Steven Bethard}}
  (Eds.). \bibinfo{publisher}{Association for Computational Linguistics},
  \bibinfo{address}{Mexico City, Mexico}, \bibinfo{pages}{2496--2506}.
\newblock
\href{https://doi.org/10.18653/v1/2024.findings-naacl.160}{doi:\nolinkurl{10.18653/v1/2024.findings-naacl.160}}


\bibitem[Deng et~al\mbox{.}(2021)]%
        {DENG2021101182}
\bibfield{author}{\bibinfo{person}{Jianfeng Deng}, \bibinfo{person}{Lianglun
  Cheng}, {and} \bibinfo{person}{Zhuowei Wang}.}
  \bibinfo{year}{2021}\natexlab{}.
\newblock \showarticletitle{Attention-based BiLSTM fused CNN with gating
  mechanism model for Chinese long text classification}.
\newblock \bibinfo{journal}{\emph{Computer Speech \& Language}}
  \bibinfo{volume}{68} (\bibinfo{year}{2021}), \bibinfo{pages}{101182}.
\newblock
\showISSN{0885-2308}
\href{https://doi.org/10.1016/j.csl.2020.101182}{doi:\nolinkurl{10.1016/j.csl.2020.101182}}


\bibitem[Du et~al\mbox{.}(2024)]%
        {10832250}
\bibfield{author}{\bibinfo{person}{Jiawei Du}, \bibinfo{person}{I-Ming Lin},
  \bibinfo{person}{I-Hsiang Chiu}, \bibinfo{person}{Xuanjun Chen},
  \bibinfo{person}{Haibin Wu}, \bibinfo{person}{Wenze Ren}, \bibinfo{person}{Yu
  Tsao}, \bibinfo{person}{Hung-Yi Lee}, {and} \bibinfo{person}{Jyh-Shing~Roger
  Jang}.} \bibinfo{year}{2024}\natexlab{}.
\newblock \showarticletitle{DFADD: The Diffusion and Flow-Matching Based Audio
  Deepfake Dataset}. In \bibinfo{booktitle}{\emph{2024 IEEE Spoken Language
  Technology Workshop (SLT)}}. \bibinfo{pages}{921--928}.
\newblock
\href{https://doi.org/10.1109/SLT61566.2024.10832250}{doi:\nolinkurl{10.1109/SLT61566.2024.10832250}}


\bibitem[Gales et~al\mbox{.}(2014)]%
        {DBLP:conf/sltu/GalesKRR14}
\bibfield{author}{\bibinfo{person}{Mark J.~F. Gales}, \bibinfo{person}{Kate~M.
  Knill}, \bibinfo{person}{Anton Ragni}, {and} \bibinfo{person}{Shakti~P.
  Rath}.} \bibinfo{year}{2014}\natexlab{}.
\newblock \showarticletitle{Speech recognition and keyword spotting for
  low-resource languages: Babel project research at {CUED}}. In
  \bibinfo{booktitle}{\emph{4th Workshop on Spoken Language Technologies for
  Under-resourced Languages, {SLTU} 2014, St. Petersburg, Russia, May 14-16,
  2014}}. \bibinfo{publisher}{{ISCA}}, \bibinfo{pages}{16--23}.
\newblock
\urldef\tempurl%
\url{https://www.isca-archive.org/sltu\_2014/gales14\_sltu.html}
\showURL{%
\tempurl}


\bibitem[Gu and Dao(2024)]%
        {gu2024mamba}
\bibfield{author}{\bibinfo{person}{Albert Gu} {and} \bibinfo{person}{Tri Dao}.}
  \bibinfo{year}{2024}\natexlab{}.
\newblock \showarticletitle{Mamba: Linear-Time Sequence Modeling with Selective
  State Spaces}. In \bibinfo{booktitle}{\emph{First Conf. on Lang. Modeling}}.
\newblock


\bibitem[Gulati et~al\mbox{.}(2020)]%
        {gulati20_interspeech}
\bibfield{author}{\bibinfo{person}{Anmol Gulati}, \bibinfo{person}{James Qin},
  \bibinfo{person}{Chung-Cheng Chiu}, \bibinfo{person}{Niki Parmar},
  \bibinfo{person}{Yu Zhang}, \bibinfo{person}{Jiahui Yu}, \bibinfo{person}{Wei
  Han}, \bibinfo{person}{Shibo Wang}, \bibinfo{person}{Zhengdong Zhang},
  \bibinfo{person}{Yonghui Wu}, {and} \bibinfo{person}{Ruoming Pang}.}
  \bibinfo{year}{2020}\natexlab{}.
\newblock \showarticletitle{Conformer: Convolution-augmented Transformer for
  Speech Recognition}. In \bibinfo{booktitle}{\emph{Interspeech 2020}}.
  \bibinfo{pages}{5036--5040}.
\newblock
\showISSN{2958-1796}
\href{https://doi.org/10.21437/Interspeech.2020-3015}{doi:\nolinkurl{10.21437/Interspeech.2020-3015}}


\bibitem[Guo et~al\mbox{.}(2024)]%
        {10447923}
\bibfield{author}{\bibinfo{person}{Yinlin Guo}, \bibinfo{person}{Haofan Huang},
  \bibinfo{person}{Xi Chen}, \bibinfo{person}{He Zhao}, {and}
  \bibinfo{person}{Yuehai Wang}.} \bibinfo{year}{2024}\natexlab{}.
\newblock \showarticletitle{Audio Deepfake Detection With Self-Supervised Wavlm
  And Multi-Fusion Attentive Classifier}. In
  \bibinfo{booktitle}{\emph{ICASSP}}.
\newblock
\href{https://doi.org/10.1109/ICASSP48485.2024.10447923}{doi:\nolinkurl{10.1109/ICASSP48485.2024.10447923}}


\bibitem[Hao et~al\mbox{.}(2025)]%
        {10889337}
\bibfield{author}{\bibinfo{person}{Yunqi Hao}, \bibinfo{person}{Minqiang Xu},
  \bibinfo{person}{Yihao Chen}, \bibinfo{person}{Yanyan Liu},
  \bibinfo{person}{Liang He}, \bibinfo{person}{Lei Fang}, {and}
  \bibinfo{person}{Lin Liu}.} \bibinfo{year}{2025}\natexlab{}.
\newblock \showarticletitle{Integrating Spectro-Temporal Cross Aggregation and
  Multi-Scale Dynamic Learning for Audio Deepfake Detection}. In
  \bibinfo{booktitle}{\emph{ICASSP 2025 - 2025 IEEE International Conference on
  Acoustics, Speech and Signal Processing (ICASSP)}}. \bibinfo{pages}{1--5}.
\newblock
\href{https://doi.org/10.1109/ICASSP49660.2025.10889337}{doi:\nolinkurl{10.1109/ICASSP49660.2025.10889337}}


\bibitem[Hsu et~al\mbox{.}(2021)]%
        {10.1109/TASLP.2021.3122291}
\bibfield{author}{\bibinfo{person}{Wei-Ning Hsu}, \bibinfo{person}{Benjamin
  Bolte}, \bibinfo{person}{Yao-Hung~Hubert Tsai}, \bibinfo{person}{Kushal
  Lakhotia}, \bibinfo{person}{Ruslan Salakhutdinov}, {and}
  \bibinfo{person}{Abdelrahman Mohamed}.} \bibinfo{year}{2021}\natexlab{}.
\newblock \showarticletitle{HuBERT: Self-Supervised Speech Representation
  Learning by Masked Prediction of Hidden Units}.
\newblock \bibinfo{journal}{\emph{IEEE/ACM Tr. ASLP}}  \bibinfo{volume}{29}
  (\bibinfo{date}{Oct.} \bibinfo{year}{2021}).
\newblock
\showISSN{2329-9290}
\href{https://doi.org/10.1109/TASLP.2021.3122291}{doi:\nolinkurl{10.1109/TASLP.2021.3122291}}


\bibitem[Hua et~al\mbox{.}(2019)]%
        {NEURIPS2019_68b1fbe7}
\bibfield{author}{\bibinfo{person}{Weizhe Hua}, \bibinfo{person}{Yuan Zhou},
  \bibinfo{person}{Christopher~M De~Sa}, \bibinfo{person}{Zhiru Zhang}, {and}
  \bibinfo{person}{G.~Edward Suh}.} \bibinfo{year}{2019}\natexlab{}.
\newblock \showarticletitle{Channel Gating Neural Networks}. In
  \bibinfo{booktitle}{\emph{Advances in Neural Information Processing
  Systems}}, \bibfield{editor}{\bibinfo{person}{H.~Wallach},
  \bibinfo{person}{H.~Larochelle}, \bibinfo{person}{A.~Beygelzimer},
  \bibinfo{person}{F.~d\textquotesingle Alch\'{e}-Buc},
  \bibinfo{person}{E.~Fox}, {and} \bibinfo{person}{R.~Garnett}} (Eds.),
  Vol.~\bibinfo{volume}{32}. \bibinfo{publisher}{Curran Associates, Inc.}
\newblock
\urldef\tempurl%
\url{https://proceedings.neurips.cc/paper_files/paper/2019/file/68b1fbe7f16e4ae3024973f12f3cb313-Paper.pdf}
\showURL{%
\tempurl}


\bibitem[Huang et~al\mbox{.}(2025)]%
        {10888328}
\bibfield{author}{\bibinfo{person}{Wen Huang}, \bibinfo{person}{Yanmei Gu},
  \bibinfo{person}{Zhiming Wang}, \bibinfo{person}{Huijia Zhu}, {and}
  \bibinfo{person}{Yanmin Qian}.} \bibinfo{year}{2025}\natexlab{}.
\newblock \showarticletitle{Generalizable Audio Deepfake Detection via Latent
  Space Refinement and Augmentation}. In \bibinfo{booktitle}{\emph{ICASSP 2025
  - 2025 IEEE International Conference on Acoustics, Speech and Signal
  Processing (ICASSP)}}. \bibinfo{pages}{1--5}.
\newblock
\href{https://doi.org/10.1109/ICASSP49660.2025.10888328}{doi:\nolinkurl{10.1109/ICASSP49660.2025.10888328}}


\bibitem[India et~al\mbox{.}(2019)]%
        {india19_interspeech}
\bibfield{author}{\bibinfo{person}{Miquel India}, \bibinfo{person}{Pooyan
  Safari}, {and} \bibinfo{person}{Javier Hernando}.}
  \bibinfo{year}{2019}\natexlab{}.
\newblock \showarticletitle{Self Multi-Head Attention for Speaker Recognition}.
  In \bibinfo{booktitle}{\emph{Interspeech}}.
\newblock
\showISSN{2958-1796}
\href{https://doi.org/10.21437/Interspeech.2019-2616}{doi:\nolinkurl{10.21437/Interspeech.2019-2616}}


\bibitem[Jiang et~al\mbox{.}(2025)]%
        {jiang2025tracing}
\bibfield{author}{\bibinfo{person}{Jiachen Jiang}, \bibinfo{person}{Jinxin
  Zhou}, {and} \bibinfo{person}{Zhihui Zhu}.} \bibinfo{year}{2025}\natexlab{}.
\newblock \showarticletitle{Tracing Representation Progression: Analyzing and
  Enhancing Layer-Wise Similarity}. In \bibinfo{booktitle}{\emph{The Thirteenth
  International Conference on Learning Representations}}.
\newblock
\urldef\tempurl%
\url{https://openreview.net/forum?id=vVxeFSR4fU}
\showURL{%
\tempurl}


\bibitem[Jin et~al\mbox{.}(2025)]%
        {10890563}
\bibfield{author}{\bibinfo{person}{Zehui Jin}, \bibinfo{person}{Linlong Lang},
  {and} \bibinfo{person}{Biao Leng}.} \bibinfo{year}{2025}\natexlab{}.
\newblock \showarticletitle{Wave-Spectrogram Cross-Modal Aggregation for Audio
  Deepfake Detection}. In \bibinfo{booktitle}{\emph{ICASSP 2025 - 2025 IEEE
  International Conference on Acoustics, Speech and Signal Processing
  (ICASSP)}}. \bibinfo{pages}{1--5}.
\newblock
\href{https://doi.org/10.1109/ICASSP49660.2025.10890563}{doi:\nolinkurl{10.1109/ICASSP49660.2025.10890563}}


\bibitem[Jung et~al\mbox{.}(2022)]%
        {9747766}
\bibfield{author}{\bibinfo{person}{Jee-weon Jung}, \bibinfo{person}{Hee-Soo
  Heo}, \bibinfo{person}{Hemlata Tak}, \bibinfo{person}{Hye-jin Shim},
  \bibinfo{person}{Joon~Son Chung}, \bibinfo{person}{Bong-Jin Lee},
  \bibinfo{person}{Ha-Jin Yu}, {and} \bibinfo{person}{Nicholas Evans}.}
  \bibinfo{year}{2022}\natexlab{}.
\newblock \showarticletitle{AASIST: Audio Anti-Spoofing Using Integrated
  Spectro-Temporal Graph Attention Networks}. In
  \bibinfo{booktitle}{\emph{ICASSP}}.
\newblock
\href{https://doi.org/10.1109/ICASSP43922.2022.9747766}{doi:\nolinkurl{10.1109/ICASSP43922.2022.9747766}}


\bibitem[Kheir et~al\mbox{.}(2025)]%
        {kheir2025comprehensive}
\bibfield{author}{\bibinfo{person}{Yassine~El Kheir}, \bibinfo{person}{Youness
  Samih}, \bibinfo{person}{Suraj Maharjan}, \bibinfo{person}{Tim Polzehl},
  {and} \bibinfo{person}{Sebastian M{\"o}ller}.}
  \bibinfo{year}{2025}\natexlab{}.
\newblock \showarticletitle{Comprehensive Layer-wise Analysis of SSL Models for
  Audio Deepfake Detection}.
\newblock \bibinfo{journal}{\emph{arXiv preprint arXiv:2502.03559}}
  (\bibinfo{year}{2025}).
\newblock


\bibitem[Kinnunen et~al\mbox{.}(2017)]%
        {kinnunen17_interspeech}
\bibfield{author}{\bibinfo{person}{Tomi Kinnunen}, \bibinfo{person}{Md.
  Sahidullah}, \bibinfo{person}{Héctor Delgado}, \bibinfo{person}{Massimiliano
  Todisco}, \bibinfo{person}{Nicholas Evans}, \bibinfo{person}{Junichi
  Yamagishi}, {and} \bibinfo{person}{Kong~Aik Lee}.}
  \bibinfo{year}{2017}\natexlab{}.
\newblock \showarticletitle{The ASVspoof 2017 Challenge: Assessing the Limits
  of Replay Spoofing Attack Detection}. In
  \bibinfo{booktitle}{\emph{Interspeech 2017}}. \bibinfo{pages}{2--6}.
\newblock
\showISSN{2958-1796}
\href{https://doi.org/10.21437/Interspeech.2017-1111}{doi:\nolinkurl{10.21437/Interspeech.2017-1111}}


\bibitem[Kornblith et~al\mbox{.}(2019)]%
        {kornblith2019similarity}
\bibfield{author}{\bibinfo{person}{Simon Kornblith}, \bibinfo{person}{Mohammad
  Norouzi}, \bibinfo{person}{Honglak Lee}, {and} \bibinfo{person}{Geoffrey
  Hinton}.} \bibinfo{year}{2019}\natexlab{}.
\newblock \showarticletitle{Similarity of neural network representations
  revisited}. In \bibinfo{booktitle}{\emph{International conference on machine
  learning}}. PMLR, \bibinfo{pages}{3519--3529}.
\newblock


\bibitem[Liu et~al\mbox{.}(2021)]%
        {liu2021pay}
\bibfield{author}{\bibinfo{person}{Hanxiao Liu}, \bibinfo{person}{Zihang Dai},
  \bibinfo{person}{David So}, {and} \bibinfo{person}{Quoc~V Le}.}
  \bibinfo{year}{2021}\natexlab{}.
\newblock \showarticletitle{Pay Attention to {MLP}s}. In
  \bibinfo{booktitle}{\emph{Advances in NeurIPS}}.
\newblock


\bibitem[Liu et~al\mbox{.}(2023)]%
        {10155166}
\bibfield{author}{\bibinfo{person}{Xuechen Liu}, \bibinfo{person}{Xin Wang},
  \bibinfo{person}{Md Sahidullah}, \bibinfo{person}{Jose Patino},
  \bibinfo{person}{Héctor Delgado}, \bibinfo{person}{Tomi Kinnunen},
  \bibinfo{person}{Massimiliano Todisco}, \bibinfo{person}{Junichi Yamagishi},
  \bibinfo{person}{Nicholas Evans}, \bibinfo{person}{Andreas Nautsch}, {and}
  \bibinfo{person}{Kong~Aik Lee}.} \bibinfo{year}{2023}\natexlab{}.
\newblock \showarticletitle{ASVspoof 2021: Towards Spoofed and Deepfake Speech
  Detection in the Wild}.
\newblock \bibinfo{journal}{\emph{IEEE/ACM Transactions on Audio, Speech, and
  Language Processing}}  \bibinfo{volume}{31} (\bibinfo{year}{2023}),
  \bibinfo{pages}{2507--2522}.
\newblock
\href{https://doi.org/10.1109/TASLP.2023.3285283}{doi:\nolinkurl{10.1109/TASLP.2023.3285283}}


\bibitem[Lorenzo-Trueba et~al\mbox{.}(2018)]%
        {lorenzotrueba18_odyssey}
\bibfield{author}{\bibinfo{person}{Jaime Lorenzo-Trueba},
  \bibinfo{person}{Junichi Yamagishi}, \bibinfo{person}{Tomoki Toda},
  \bibinfo{person}{Daisuke Saito}, \bibinfo{person}{Fernando Villavicencio},
  \bibinfo{person}{Tomi Kinnunen}, {and} \bibinfo{person}{Zhenhua Ling}.}
  \bibinfo{year}{2018}\natexlab{}.
\newblock \showarticletitle{The Voice Conversion Challenge 2018: Promoting
  Development of Parallel and Nonparallel Methods}. In
  \bibinfo{booktitle}{\emph{The Speaker and Language Recognition Workshop
  (Odyssey 2018)}}. \bibinfo{pages}{195--202}.
\newblock
\href{https://doi.org/10.21437/Odyssey.2018-28}{doi:\nolinkurl{10.21437/Odyssey.2018-28}}


\bibitem[Marek et~al\mbox{.}(2024)]%
        {marek2024audio}
\bibfield{author}{\bibinfo{person}{Bart{\l}omiej Marek}, \bibinfo{person}{Piotr
  Kawa}, {and} \bibinfo{person}{Piotr Syga}.} \bibinfo{year}{2024}\natexlab{}.
\newblock \showarticletitle{Are audio DeepFake detection models polyglots?}
\newblock \bibinfo{journal}{\emph{arXiv preprint arXiv:2412.17924}}
  (\bibinfo{year}{2024}).
\newblock


\bibitem[Martín-Doñas and Álvarez(2022)]%
        {9747768}
\bibfield{author}{\bibinfo{person}{Juan~M. Martín-Doñas} {and}
  \bibinfo{person}{Aitor Álvarez}.} \bibinfo{year}{2022}\natexlab{}.
\newblock \showarticletitle{The Vicomtech Audio Deepfake Detection System Based
  on Wav2vec2 for the 2022 ADD Challenge}. In \bibinfo{booktitle}{\emph{ICASSP
  2022 - 2022 IEEE International Conference on Acoustics, Speech and Signal
  Processing (ICASSP)}}. \bibinfo{pages}{9241--9245}.
\newblock
\href{https://doi.org/10.1109/ICASSP43922.2022.9747768}{doi:\nolinkurl{10.1109/ICASSP43922.2022.9747768}}


\bibitem[Müller et~al\mbox{.}(2022)]%
        {muller22_interspeech}
\bibfield{author}{\bibinfo{person}{Nicolas Müller}, \bibinfo{person}{Pavel
  Czempin}, \bibinfo{person}{Franziska Diekmann}, \bibinfo{person}{Adam
  Froghyar}, {and} \bibinfo{person}{Konstantin Böttinger}.}
  \bibinfo{year}{2022}\natexlab{}.
\newblock \showarticletitle{Does Audio Deepfake Detection Generalize?}. In
  \bibinfo{booktitle}{\emph{Interspeech}}.
\newblock
\showISSN{2958-1796}
\href{https://doi.org/10.21437/Interspeech.2022-108}{doi:\nolinkurl{10.21437/Interspeech.2022-108}}


\bibitem[Müller et~al\mbox{.}(2024)]%
        {10650962}
\bibfield{author}{\bibinfo{person}{Nicolas~M. Müller}, \bibinfo{person}{Piotr
  Kawa}, \bibinfo{person}{Wei~Herng Choong}, \bibinfo{person}{Edresson
  Casanova}, \bibinfo{person}{Eren Gölge}, \bibinfo{person}{Thorsten Müller},
  \bibinfo{person}{Piotr Syga}, \bibinfo{person}{Philip Sperl}, {and}
  \bibinfo{person}{Konstantin Böttinger}.} \bibinfo{year}{2024}\natexlab{}.
\newblock \showarticletitle{MLAAD: The Multi-Language Audio Anti-Spoofing
  Dataset}. In \bibinfo{booktitle}{\emph{2024 International Joint Conference on
  Neural Networks (IJCNN)}}. \bibinfo{pages}{1--7}.
\newblock
\href{https://doi.org/10.1109/IJCNN60899.2024.10650962}{doi:\nolinkurl{10.1109/IJCNN60899.2024.10650962}}


\bibitem[Pan et~al\mbox{.}(2024)]%
        {pan24c_interspeech}
\bibfield{author}{\bibinfo{person}{Zihan Pan}, \bibinfo{person}{Tianchi Liu},
  \bibinfo{person}{Hardik~B. Sailor}, {and} \bibinfo{person}{Qiongqiong Wang}.}
  \bibinfo{year}{2024}\natexlab{}.
\newblock \showarticletitle{Attentive Merging of Hidden Embeddings from
  Pre-trained Speech Model for Anti-spoofing Detection}. In
  \bibinfo{booktitle}{\emph{Interspeech}}.
\newblock
\showISSN{2958-1796}
\href{https://doi.org/10.21437/Interspeech.2024-1472}{doi:\nolinkurl{10.21437/Interspeech.2024-1472}}


\bibitem[Papyan et~al\mbox{.}(2020)]%
        {doi:10.1073/pnas.2015509117}
\bibfield{author}{\bibinfo{person}{Vardan Papyan}, \bibinfo{person}{X.~Y. Han},
  {and} \bibinfo{person}{David~L. Donoho}.} \bibinfo{year}{2020}\natexlab{}.
\newblock \showarticletitle{Prevalence of neural collapse during the terminal
  phase of deep learning training}.
\newblock \bibinfo{journal}{\emph{Proceedings of the National Academy of
  Sciences}} \bibinfo{volume}{117}, \bibinfo{number}{40}
  (\bibinfo{year}{2020}), \bibinfo{pages}{24652--24663}.
\newblock
\showeprint{https://www.pnas.org/doi/pdf/10.1073/pnas.2015509117}
\href{https://doi.org/10.1073/pnas.2015509117}{doi:\nolinkurl{10.1073/pnas.2015509117}}


\bibitem[Pasad et~al\mbox{.}(2021)]%
        {9688093}
\bibfield{author}{\bibinfo{person}{Ankita Pasad}, \bibinfo{person}{Ju-Chieh
  Chou}, {and} \bibinfo{person}{Karen Livescu}.}
  \bibinfo{year}{2021}\natexlab{}.
\newblock \showarticletitle{Layer-Wise Analysis of a Self-Supervised Speech
  Representation Model}. In \bibinfo{booktitle}{\emph{2021 IEEE Automatic
  Speech Recognition and Understanding Workshop (ASRU)}}.
  \bibinfo{pages}{914--921}.
\newblock
\href{https://doi.org/10.1109/ASRU51503.2021.9688093}{doi:\nolinkurl{10.1109/ASRU51503.2021.9688093}}


\bibitem[Pasad et~al\mbox{.}(2023)]%
        {10096149}
\bibfield{author}{\bibinfo{person}{Ankita Pasad}, \bibinfo{person}{Bowen Shi},
  {and} \bibinfo{person}{Karen Livescu}.} \bibinfo{year}{2023}\natexlab{}.
\newblock \showarticletitle{Comparative Layer-Wise Analysis of Self-Supervised
  Speech Models}. In \bibinfo{booktitle}{\emph{ICASSP 2023 - 2023 IEEE
  International Conference on Acoustics, Speech and Signal Processing
  (ICASSP)}}. \bibinfo{pages}{1--5}.
\newblock
\href{https://doi.org/10.1109/ICASSP49357.2023.10096149}{doi:\nolinkurl{10.1109/ICASSP49357.2023.10096149}}


\bibitem[Prabhu et~al\mbox{.}(2024)]%
        {prabhu24_interspeech}
\bibfield{author}{\bibinfo{person}{Darshan Prabhu}, \bibinfo{person}{Yifan
  Peng}, \bibinfo{person}{Preethi Jyothi}, {and} \bibinfo{person}{Shinji
  Watanabe}.} \bibinfo{year}{2024}\natexlab{}.
\newblock \showarticletitle{MULTI-CONVFORMER: Extending Conformer with Multiple
  Convolution Kernels}. In \bibinfo{booktitle}{\emph{Interspeech}}.
\newblock
\showISSN{2958-1796}
\href{https://doi.org/10.21437/Interspeech.2024-2384}{doi:\nolinkurl{10.21437/Interspeech.2024-2384}}


\bibitem[Pratap et~al\mbox{.}(2024)]%
        {JMLR:v25:23-1318}
\bibfield{author}{\bibinfo{person}{Vineel Pratap}, \bibinfo{person}{Andros
  Tjandra}, \bibinfo{person}{Bowen Shi}, \bibinfo{person}{Paden Tomasello},
  \bibinfo{person}{Arun Babu}, \bibinfo{person}{Sayani Kundu},
  \bibinfo{person}{Ali Elkahky}, \bibinfo{person}{Zhaoheng Ni},
  \bibinfo{person}{Apoorv Vyas}, \bibinfo{person}{Maryam Fazel-Zarandi},
  \bibinfo{person}{Alexei Baevski}, \bibinfo{person}{Yossi Adi},
  \bibinfo{person}{Xiaohui Zhang}, \bibinfo{person}{Wei-Ning Hsu},
  \bibinfo{person}{Alexis Conneau}, {and} \bibinfo{person}{Michael Auli}.}
  \bibinfo{year}{2024}\natexlab{}.
\newblock \showarticletitle{Scaling Speech Technology to 1,000+ Languages}.
\newblock \bibinfo{journal}{\emph{Journal of Machine Learning Research}}
  \bibinfo{volume}{25}, \bibinfo{number}{97} (\bibinfo{year}{2024}).
\newblock


\bibitem[Pratap et~al\mbox{.}(2020)]%
        {pratap20_interspeech}
\bibfield{author}{\bibinfo{person}{Vineel Pratap}, \bibinfo{person}{Qiantong
  Xu}, \bibinfo{person}{Anuroop Sriram}, \bibinfo{person}{Gabriel Synnaeve},
  {and} \bibinfo{person}{Ronan Collobert}.} \bibinfo{year}{2020}\natexlab{}.
\newblock \showarticletitle{MLS: A Large-Scale Multilingual Dataset for Speech
  Research}. In \bibinfo{booktitle}{\emph{Interspeech 2020}}.
  \bibinfo{pages}{2757--2761}.
\newblock
\showISSN{2958-1796}
\href{https://doi.org/10.21437/Interspeech.2020-2826}{doi:\nolinkurl{10.21437/Interspeech.2020-2826}}


\bibitem[Reimao and Tzerpos(2019)]%
        {8906599}
\bibfield{author}{\bibinfo{person}{Ricardo Reimao} {and}
  \bibinfo{person}{Vassilios Tzerpos}.} \bibinfo{year}{2019}\natexlab{}.
\newblock \showarticletitle{FoR: A Dataset for Synthetic Speech Detection}. In
  \bibinfo{booktitle}{\emph{2019 International Conference on Speech Technology
  and Human-Computer Dialogue (SpeD)}}. \bibinfo{pages}{1--10}.
\newblock
\href{https://doi.org/10.1109/SPED.2019.8906599}{doi:\nolinkurl{10.1109/SPED.2019.8906599}}


\bibitem[Rosello et~al\mbox{.}(2023)]%
        {rosello23_interspeech}
\bibfield{author}{\bibinfo{person}{Eros Rosello}, \bibinfo{person}{Alejandro
  Gomez-Alanis}, \bibinfo{person}{Angel~M. Gomez}, {and}
  \bibinfo{person}{Antonio Peinado}.} \bibinfo{year}{2023}\natexlab{}.
\newblock \showarticletitle{A conformer-based classifier for variable-length
  utterance processing in anti-spoofing}. In
  \bibinfo{booktitle}{\emph{Interspeech}}.
\newblock
\showISSN{2958-1796}
\href{https://doi.org/10.21437/Interspeech.2023-1820}{doi:\nolinkurl{10.21437/Interspeech.2023-1820}}


\bibitem[Shazeer(2020)]%
        {shazeer2020glu}
\bibfield{author}{\bibinfo{person}{Noam Shazeer}.}
  \bibinfo{year}{2020}\natexlab{}.
\newblock \showarticletitle{Glu variants improve transformer}.
\newblock \bibinfo{journal}{\emph{arXiv preprint arXiv:2002.05202}}
  (\bibinfo{year}{2020}).
\newblock


\bibitem[Sun et~al\mbox{.}(2023)]%
        {Sun_2023_CVPR}
\bibfield{author}{\bibinfo{person}{Chengzhe Sun}, \bibinfo{person}{Shan Jia},
  \bibinfo{person}{Shuwei Hou}, {and} \bibinfo{person}{Siwei Lyu}.}
  \bibinfo{year}{2023}\natexlab{}.
\newblock \showarticletitle{AI-Synthesized Voice Detection Using Neural Vocoder
  Artifacts}. In \bibinfo{booktitle}{\emph{Proceedings of the IEEE/CVF
  Conference on Computer Vision and Pattern Recognition (CVPR) Workshops}}.
  \bibinfo{pages}{904--912}.
\newblock


\bibitem[Tak et~al\mbox{.}(2022a)]%
        {9746213}
\bibfield{author}{\bibinfo{person}{Hemlata Tak}, \bibinfo{person}{Madhu
  Kamble}, \bibinfo{person}{Jose Patino}, \bibinfo{person}{Massimiliano
  Todisco}, {and} \bibinfo{person}{Nicholas Evans}.}
  \bibinfo{year}{2022}\natexlab{a}.
\newblock \showarticletitle{Rawboost: A Raw Data Boosting and Augmentation
  Method Applied to Automatic Speaker Verification Anti-Spoofing}. In
  \bibinfo{booktitle}{\emph{ICASSP}}.
\newblock
\href{https://doi.org/10.1109/ICASSP43922.2022.9746213}{doi:\nolinkurl{10.1109/ICASSP43922.2022.9746213}}


\bibitem[Tak et~al\mbox{.}(2022b)]%
        {tak22_odyssey}
\bibfield{author}{\bibinfo{person}{Hemlata Tak}, \bibinfo{person}{Massimiliano
  Todisco}, \bibinfo{person}{Xin Wang}, \bibinfo{person}{Jee weon Jung},
  \bibinfo{person}{Junichi Yamagishi}, {and} \bibinfo{person}{Nicholas Evans}.}
  \bibinfo{year}{2022}\natexlab{b}.
\newblock \showarticletitle{Automatic Speaker Verification Spoofing and
  Deepfake Detection Using Wav2vec 2.0 and Data Augmentation}. In
  \bibinfo{booktitle}{\emph{The SLR Workshop (Odyssey 2022)}}.
\newblock
\href{https://doi.org/10.21437/Odyssey.2022-16}{doi:\nolinkurl{10.21437/Odyssey.2022-16}}


\bibitem[{Tamayo Flórez} et~al\mbox{.}(2023)]%
        {tamayoflorez23_interspeech}
\bibfield{author}{\bibinfo{person}{Pablo~Andrés {Tamayo Flórez}},
  \bibinfo{person}{Rubén Manrique}, {and} \bibinfo{person}{Bernardo {Pereira
  Nunes}}.} \bibinfo{year}{2023}\natexlab{}.
\newblock \showarticletitle{HABLA: A Dataset of Latin American Spanish Accents
  for Voice Anti-spoofing}. In \bibinfo{booktitle}{\emph{Interspeech 2023}}.
  \bibinfo{pages}{1963--1967}.
\newblock
\showISSN{2958-1796}
\href{https://doi.org/10.21437/Interspeech.2023-2272}{doi:\nolinkurl{10.21437/Interspeech.2023-2272}}


\bibitem[Todisco et~al\mbox{.}(2017)]%
        {TODISCO2017516}
\bibfield{author}{\bibinfo{person}{Massimiliano Todisco},
  \bibinfo{person}{Héctor Delgado}, {and} \bibinfo{person}{Nicholas Evans}.}
  \bibinfo{year}{2017}\natexlab{}.
\newblock \showarticletitle{Constant Q cepstral coefficients: A spoofing
  countermeasure for automatic speaker verification}.
\newblock \bibinfo{journal}{\emph{Computer Speech \& Language}}
  \bibinfo{volume}{45} (\bibinfo{year}{2017}), \bibinfo{pages}{516--535}.
\newblock
\showISSN{0885-2308}
\href{https://doi.org/10.1016/j.csl.2017.01.001}{doi:\nolinkurl{10.1016/j.csl.2017.01.001}}


\bibitem[Todisco et~al\mbox{.}(2019)]%
        {todisco2019asvspoof}
\bibfield{author}{\bibinfo{person}{Massimiliano Todisco}, \bibinfo{person}{Xin
  Wang}, \bibinfo{person}{Ville Vestman}, \bibinfo{person}{Md Sahidullah},
  \bibinfo{person}{H{\'e}ctor Delgado}, \bibinfo{person}{Andreas Nautsch},
  \bibinfo{person}{Junichi Yamagishi}, \bibinfo{person}{Nicholas Evans},
  \bibinfo{person}{Tomi~H Kinnunen}, {and} \bibinfo{person}{Kong~Aik Lee}.}
  \bibinfo{year}{2019}\natexlab{}.
\newblock \showarticletitle{{ASV}spoof 2019: Future Horizons in Spoofed and
  Fake Audio Detection}. In \bibinfo{booktitle}{\emph{Interspeech}}.
\newblock


\bibitem[Tran et~al\mbox{.}(2024)]%
        {tran24_interspeech}
\bibfield{author}{\bibinfo{person}{Hoan~My Tran}, \bibinfo{person}{David
  Guennec}, \bibinfo{person}{Philippe Martin}, \bibinfo{person}{Aghilas Sini},
  \bibinfo{person}{Damien Lolive}, \bibinfo{person}{Arnaud Delhay}, {and}
  \bibinfo{person}{Pierre-François Marteau}.} \bibinfo{year}{2024}\natexlab{}.
\newblock \showarticletitle{Spoofed Speech Detection with a Focus on Speaker
  Embedding}. In \bibinfo{booktitle}{\emph{Interspeech}}.
\newblock
\showISSN{2958-1796}
\href{https://doi.org/10.21437/Interspeech.2024-481}{doi:\nolinkurl{10.21437/Interspeech.2024-481}}


\bibitem[Truong et~al\mbox{.}(2024)]%
        {truong24b_interspeech}
\bibfield{author}{\bibinfo{person}{Duc-Tuan Truong}, \bibinfo{person}{Ruijie
  Tao}, \bibinfo{person}{Tuan Nguyen}, \bibinfo{person}{Hieu-Thi Luong},
  \bibinfo{person}{Kong~Aik Lee}, {and} \bibinfo{person}{Eng~Siong Chng}.}
  \bibinfo{year}{2024}\natexlab{}.
\newblock \showarticletitle{Temporal-Channel Modeling in Multi-head
  Self-Attention for Synthetic Speech Detection}. In
  \bibinfo{booktitle}{\emph{Interspeech}}.
\newblock
\showISSN{2958-1796}
\href{https://doi.org/10.21437/Interspeech.2024-659}{doi:\nolinkurl{10.21437/Interspeech.2024-659}}


\bibitem[Valk and Alum{\"a}e(2021)]%
        {valk2021slt}
\bibfield{author}{\bibinfo{person}{J{\"o}rgen Valk} {and}
  \bibinfo{person}{Tanel Alum{\"a}e}.} \bibinfo{year}{2021}\natexlab{}.
\newblock \showarticletitle{{VoxLingua107}: a Dataset for Spoken Language
  Recognition}. In \bibinfo{booktitle}{\emph{Proc. IEEE SLT Workshop}}.
\newblock


\bibitem[Vaswani et~al\mbox{.}(2017)]%
        {NIPS2017_3f5ee243}
\bibfield{author}{\bibinfo{person}{Ashish Vaswani}, \bibinfo{person}{Noam
  Shazeer}, \bibinfo{person}{Niki Parmar}, \bibinfo{person}{Jakob Uszkoreit},
  \bibinfo{person}{Llion Jones}, \bibinfo{person}{Aidan~N Gomez},
  \bibinfo{person}{\L~ukasz Kaiser}, {and} \bibinfo{person}{Illia Polosukhin}.}
  \bibinfo{year}{2017}\natexlab{}.
\newblock \showarticletitle{Attention is All you Need}. In
  \bibinfo{booktitle}{\emph{Advances in NeurIPS}}, Vol.~\bibinfo{volume}{30}.
\newblock


\bibitem[Wang et~al\mbox{.}(2024c)]%
        {10658409}
\bibfield{author}{\bibinfo{person}{Bor-Shiun Wang}, \bibinfo{person}{Chien-Yi
  Wang}, {and} \bibinfo{person}{Wei-Chen Chiu}.}
  \bibinfo{year}{2024}\natexlab{c}.
\newblock \showarticletitle{MCPNet: An Interpretable Classifier via Multi-Level
  Concept Prototypes}. In \bibinfo{booktitle}{\emph{2024 IEEE/CVF Conference on
  Computer Vision and Pattern Recognition (CVPR)}}.
  \bibinfo{pages}{10885--10894}.
\newblock
\href{https://doi.org/10.1109/CVPR52733.2024.01035}{doi:\nolinkurl{10.1109/CVPR52733.2024.01035}}


\bibitem[Wang et~al\mbox{.}(2021)]%
        {wang-etal-2021-voxpopuli}
\bibfield{author}{\bibinfo{person}{Changhan Wang}, \bibinfo{person}{Morgane
  Riviere}, \bibinfo{person}{Ann Lee}, \bibinfo{person}{Anne Wu},
  \bibinfo{person}{Chaitanya Talnikar}, \bibinfo{person}{Daniel Haziza},
  \bibinfo{person}{Mary Williamson}, \bibinfo{person}{Juan Pino}, {and}
  \bibinfo{person}{Emmanuel Dupoux}.} \bibinfo{year}{2021}\natexlab{}.
\newblock \showarticletitle{{V}ox{P}opuli: A Large-Scale Multilingual Speech
  Corpus for Representation Learning, Semi-Supervised Learning and
  Interpretation}. In \bibinfo{booktitle}{\emph{Proceedings of the 59th Annual
  Meeting of the Association for Computational Linguistics and the 11th
  International Joint Conference on Natural Language Processing (Volume 1: Long
  Papers)}}, \bibfield{editor}{\bibinfo{person}{Chengqing Zong},
  \bibinfo{person}{Fei Xia}, \bibinfo{person}{Wenjie Li}, {and}
  \bibinfo{person}{Roberto Navigli}} (Eds.). \bibinfo{publisher}{Association
  for Computational Linguistics}, \bibinfo{address}{Online},
  \bibinfo{pages}{993--1003}.
\newblock
\href{https://doi.org/10.18653/v1/2021.acl-long.80}{doi:\nolinkurl{10.18653/v1/2021.acl-long.80}}


\bibitem[Wang et~al\mbox{.}(2024a)]%
        {wang24_asvspoof}
\bibfield{author}{\bibinfo{person}{Xin Wang}, \bibinfo{person}{Héctor
  Delgado}, \bibinfo{person}{Hemlata Tak}, \bibinfo{person}{Jee weon Jung},
  \bibinfo{person}{Hye jin Shim}, \bibinfo{person}{Massimiliano Todisco},
  \bibinfo{person}{Ivan Kukanov}, \bibinfo{person}{Xuechen Liu},
  \bibinfo{person}{Md Sahidullah}, \bibinfo{person}{Tomi~H. Kinnunen},
  \bibinfo{person}{Nicholas Evans}, \bibinfo{person}{Kong~Aik Lee}, {and}
  \bibinfo{person}{Junichi Yamagishi}.} \bibinfo{year}{2024}\natexlab{a}.
\newblock \showarticletitle{ASVspoof 5: crowdsourced speech data, deepfakes,
  and adversarial attacks at scale}. In \bibinfo{booktitle}{\emph{ASVspoof}}.
\newblock
\href{https://doi.org/10.21437/ASVspoof.2024-1}{doi:\nolinkurl{10.21437/ASVspoof.2024-1}}


\bibitem[Wang et~al\mbox{.}(2020)]%
        {WANG2020101114}
\bibfield{author}{\bibinfo{person}{Xin Wang}, \bibinfo{person}{Junichi
  Yamagishi}, \bibinfo{person}{Massimiliano Todisco}, \bibinfo{person}{Héctor
  Delgado}, \bibinfo{person}{Andreas Nautsch}, \bibinfo{person}{Nicholas
  Evans}, \bibinfo{person}{Md Sahidullah}, \bibinfo{person}{Ville Vestman},
  \bibinfo{person}{Tomi Kinnunen}, \bibinfo{person}{Kong~Aik Lee},
  \bibinfo{person}{Lauri Juvela}, \bibinfo{person}{Paavo Alku},
  \bibinfo{person}{Yu-Huai Peng}, \bibinfo{person}{Hsin-Te Hwang},
  \bibinfo{person}{Yu Tsao}, \bibinfo{person}{Hsin-Min Wang},
  \bibinfo{person}{Sébastien~Le Maguer}, \bibinfo{person}{Markus Becker},
  \bibinfo{person}{Fergus Henderson}, \bibinfo{person}{Rob Clark},
  \bibinfo{person}{Yu Zhang}, \bibinfo{person}{Quan Wang}, \bibinfo{person}{Ye
  Jia}, \bibinfo{person}{Kai Onuma}, \bibinfo{person}{Koji Mushika},
  \bibinfo{person}{Takashi Kaneda}, \bibinfo{person}{Yuan Jiang},
  \bibinfo{person}{Li-Juan Liu}, \bibinfo{person}{Yi-Chiao Wu},
  \bibinfo{person}{Wen-Chin Huang}, \bibinfo{person}{Tomoki Toda},
  \bibinfo{person}{Kou Tanaka}, \bibinfo{person}{Hirokazu Kameoka},
  \bibinfo{person}{Ingmar Steiner}, \bibinfo{person}{Driss Matrouf},
  \bibinfo{person}{Jean-François Bonastre}, \bibinfo{person}{Avashna
  Govender}, \bibinfo{person}{Srikanth Ronanki}, \bibinfo{person}{Jing-Xuan
  Zhang}, {and} \bibinfo{person}{Zhen-Hua Ling}.}
  \bibinfo{year}{2020}\natexlab{}.
\newblock \showarticletitle{ASVspoof 2019: A large-scale public database of
  synthesized, converted and replayed speech}.
\newblock \bibinfo{journal}{\emph{Computer Speech \& Language}}
  \bibinfo{volume}{64} (\bibinfo{year}{2020}), \bibinfo{pages}{101114}.
\newblock
\showISSN{0885-2308}
\href{https://doi.org/10.1016/j.csl.2020.101114}{doi:\nolinkurl{10.1016/j.csl.2020.101114}}


\bibitem[Wang et~al\mbox{.}(2024b)]%
        {wang2024mixture}
\bibfield{author}{\bibinfo{person}{Zhiyong Wang}, \bibinfo{person}{Ruibo Fu},
  \bibinfo{person}{Zhengqi Wen}, \bibinfo{person}{Jianhua Tao},
  \bibinfo{person}{Xiaopeng Wang}, \bibinfo{person}{Yuankun Xie},
  \bibinfo{person}{Xin Qi}, \bibinfo{person}{Shuchen Shi}, \bibinfo{person}{Yi
  Lu}, \bibinfo{person}{Yukun Liu}, {et~al\mbox{.}}}
  \bibinfo{year}{2024}\natexlab{b}.
\newblock \showarticletitle{Mixture of experts fusion for fake audio detection
  using frozen wav2vec 2.0}.
\newblock \bibinfo{journal}{\emph{arXiv preprint arXiv:2409.11909}}
  (\bibinfo{year}{2024}).
\newblock


\bibitem[Wu et~al\mbox{.}(2015)]%
        {wu15e_interspeech}
\bibfield{author}{\bibinfo{person}{Zhizheng Wu}, \bibinfo{person}{Tomi
  Kinnunen}, \bibinfo{person}{Nicholas Evans}, \bibinfo{person}{Junichi
  Yamagishi}, \bibinfo{person}{Cemal Hanilçi}, \bibinfo{person}{Md.
  Sahidullah}, {and} \bibinfo{person}{Aleksandr Sizov}.}
  \bibinfo{year}{2015}\natexlab{}.
\newblock \showarticletitle{ASVspoof 2015: the first automatic speaker
  verification spoofing and countermeasures challenge}. In
  \bibinfo{booktitle}{\emph{Interspeech 2015}}. \bibinfo{pages}{2037--2041}.
\newblock
\showISSN{2958-1796}
\href{https://doi.org/10.21437/Interspeech.2015-462}{doi:\nolinkurl{10.21437/Interspeech.2015-462}}


\bibitem[Xiao and Das(2025)]%
        {10909468}
\bibfield{author}{\bibinfo{person}{Yang Xiao} {and}
  \bibinfo{person}{Rohan~Kumar Das}.} \bibinfo{year}{2025}\natexlab{}.
\newblock \showarticletitle{XLSR-Mamba: A Dual-Column Bidirectional State Space
  Model for Spoofing Attack Detection}.
\newblock \bibinfo{journal}{\emph{IEEE Signal Processing Letters}}
  \bibinfo{volume}{32} (\bibinfo{year}{2025}), \bibinfo{pages}{1276--1280}.
\newblock
\href{https://doi.org/10.1109/LSP.2025.3547861}{doi:\nolinkurl{10.1109/LSP.2025.3547861}}


\bibitem[Yamagishi et~al\mbox{.}(2019)]%
        {Yamagishi2019CSTRVC}
\bibfield{author}{\bibinfo{person}{Junichi Yamagishi},
  \bibinfo{person}{Christophe Veaux}, {and} \bibinfo{person}{Kirsten
  MacDonald}.} \bibinfo{year}{2019}\natexlab{}.
\newblock \showarticletitle{CSTR VCTK Corpus: English Multi-speaker Corpus for
  CSTR Voice Cloning Toolkit}.
\newblock
\urldef\tempurl%
\url{https://api.semanticscholar.org/CorpusID:213060286}
\showURL{%
\tempurl}


\bibitem[Yamagishi et~al\mbox{.}(2021)]%
        {yamagishi21_asvspoof}
\bibfield{author}{\bibinfo{person}{Junichi Yamagishi}, \bibinfo{person}{Xin
  Wang}, \bibinfo{person}{Massimiliano Todisco}, \bibinfo{person}{Md
  Sahidullah}, \bibinfo{person}{Jose Patino}, \bibinfo{person}{Andreas
  Nautsch}, \bibinfo{person}{Xuechen Liu}, \bibinfo{person}{Kong~Aik Lee},
  \bibinfo{person}{Tomi Kinnunen}, \bibinfo{person}{Nicholas Evans}, {and}
  \bibinfo{person}{Héctor Delgado}.} \bibinfo{year}{2021}\natexlab{}.
\newblock \showarticletitle{ASVspoof 2021: accelerating progress in spoofed and
  deepfake speech detection}. In \bibinfo{booktitle}{\emph{2021 Edition of the
  Automatic Speaker Verification and Spoofing Countermeasures Challenge}}.
  \bibinfo{pages}{47--54}.
\newblock
\href{https://doi.org/10.21437/ASVSPOOF.2021-8}{doi:\nolinkurl{10.21437/ASVSPOOF.2021-8}}


\bibitem[Yi et~al\mbox{.}(2023)]%
        {DBLP:conf/dada/YiTFYWWZZZRXZGW23}
\bibfield{author}{\bibinfo{person}{Jiangyan Yi}, \bibinfo{person}{Jianhua Tao},
  \bibinfo{person}{Ruibo Fu}, \bibinfo{person}{Xinrui Yan},
  \bibinfo{person}{Chenglong Wang}, \bibinfo{person}{Tao Wang},
  \bibinfo{person}{Chu~Yuan Zhang}, \bibinfo{person}{Xiaohui Zhang},
  \bibinfo{person}{Yan Zhao}, \bibinfo{person}{Yong Ren}, \bibinfo{person}{Le
  Xu}, \bibinfo{person}{Junzuo Zhou}, \bibinfo{person}{Hao Gu},
  \bibinfo{person}{Zhengqi Wen}, \bibinfo{person}{Shan Liang},
  \bibinfo{person}{Zheng Lian}, \bibinfo{person}{Shuai Nie}, {and}
  \bibinfo{person}{Haizhou Li}.} \bibinfo{year}{2023}\natexlab{}.
\newblock \showarticletitle{{ADD} 2023: the Second Audio Deepfake Detection
  Challenge}. In \bibinfo{booktitle}{\emph{Proceedings of the Workshop on
  Deepfake Audio Detection and Analysis co-located with 32th International
  Joint Conference on Artificial Intelligence {(IJCAI} 2023), Macao, China,
  August 19, 2023}} \emph{(\bibinfo{series}{{CEUR} Workshop Proceedings},
  Vol.~\bibinfo{volume}{3597})}, \bibfield{editor}{\bibinfo{person}{Jianhua
  Tao}, \bibinfo{person}{Haizhou Li}, \bibinfo{person}{Jiangyan Yi}, {and}
  \bibinfo{person}{Cunhang Fan}} (Eds.). \bibinfo{publisher}{CEUR-WS.org},
  \bibinfo{pages}{125--130}.
\newblock
\urldef\tempurl%
\url{https://ceur-ws.org/Vol-3597/paper21.pdf}
\showURL{%
\tempurl}


\bibitem[Yi et~al\mbox{.}(2020)]%
        {yi20_vccbc}
\bibfield{author}{\bibinfo{person}{Zhao Yi}, \bibinfo{person}{Wen-Chin Huang},
  \bibinfo{person}{Xiaohai Tian}, \bibinfo{person}{Junichi Yamagishi},
  \bibinfo{person}{Rohan~Kumar Das}, \bibinfo{person}{Tomi Kinnunen},
  \bibinfo{person}{Zhen-Hua Ling}, {and} \bibinfo{person}{Tomoki Toda}.}
  \bibinfo{year}{2020}\natexlab{}.
\newblock \showarticletitle{Voice Conversion Challenge 2020 –- Intra-lingual
  semi-parallel and cross-lingual voice conversion –-}. In
  \bibinfo{booktitle}{\emph{Joint Workshop for the Blizzard Challenge and Voice
  Conversion Challenge 2020}}. \bibinfo{pages}{80--98}.
\newblock
\href{https://doi.org/10.21437/VCCBC.2020-14}{doi:\nolinkurl{10.21437/VCCBC.2020-14}}


\bibitem[Zhang et~al\mbox{.}(2024b)]%
        {zhang24j_interspeech}
\bibfield{author}{\bibinfo{person}{Lin Zhang}, \bibinfo{person}{Xin Wang},
  \bibinfo{person}{Erica Cooper}, \bibinfo{person}{Mireia Diez},
  \bibinfo{person}{Federico Landini}, \bibinfo{person}{Nicholas Evans}, {and}
  \bibinfo{person}{Junichi Yamagishi}.} \bibinfo{year}{2024}\natexlab{b}.
\newblock \showarticletitle{Spoof Diarization: "What Spoofed When" in Partially
  Spoofed Audio}. In \bibinfo{booktitle}{\emph{Interspeech}}.
\newblock
\showISSN{2958-1796}
\href{https://doi.org/10.21437/Interspeech.2024-1365}{doi:\nolinkurl{10.21437/Interspeech.2024-1365}}


\bibitem[Zhang et~al\mbox{.}(2023)]%
        {zhang23v_interspeech}
\bibfield{author}{\bibinfo{person}{Lin Zhang}, \bibinfo{person}{Xin Wang},
  \bibinfo{person}{Erica Cooper}, \bibinfo{person}{Nicholas Evans}, {and}
  \bibinfo{person}{Junichi Yamagishi}.} \bibinfo{year}{2023}\natexlab{}.
\newblock \showarticletitle{Range-Based Equal Error Rate for Spoof
  Localization}. In \bibinfo{booktitle}{\emph{Interspeech}}.
\newblock
\showISSN{2958-1796}
\href{https://doi.org/10.21437/Interspeech.2023-1214}{doi:\nolinkurl{10.21437/Interspeech.2023-1214}}


\bibitem[Zhang et~al\mbox{.}(2024c)]%
        {zhang2024audio}
\bibfield{author}{\bibinfo{person}{Qishan Zhang}, \bibinfo{person}{Shuangbing
  Wen}, {and} \bibinfo{person}{Tao Hu}.} \bibinfo{year}{2024}\natexlab{c}.
\newblock \showarticletitle{Audio Deepfake Detection with Self-Supervised
  {XLS}-R and {SLS} Classifier}. In \bibinfo{booktitle}{\emph{ACM Multimedia
  2024}}.
\newblock


\bibitem[Zhang et~al\mbox{.}(2024a)]%
        {10448049}
\bibfield{author}{\bibinfo{person}{Yuxiang Zhang}, \bibinfo{person}{Jingze Lu},
  \bibinfo{person}{Zengqiang Shang}, \bibinfo{person}{Wenchao Wang}, {and}
  \bibinfo{person}{Pengyuan Zhang}.} \bibinfo{year}{2024}\natexlab{a}.
\newblock \showarticletitle{Improving Short Utterance Anti-Spoofing with
  Aasist2}. In \bibinfo{booktitle}{\emph{ICASSP}}.
\newblock
\href{https://doi.org/10.1109/ICASSP48485.2024.10448049}{doi:\nolinkurl{10.1109/ICASSP48485.2024.10448049}}


\end{thebibliography}

\end{document}